\documentclass[12pt]{article}

\usepackage{amsmath}
\usepackage{graphicx,psfrag,epsf,subfig}
\usepackage{enumerate}
\usepackage{natbib}
\usepackage{amssymb, amsfonts, amscd, xspace, pifont, amsthm,bm,bbm}
\usepackage{mathrsfs}
\usepackage{booktabs}
\usepackage{multirow}
\usepackage{epstopdf}
\usepackage{color}
\usepackage{algorithm} 
\usepackage{algpseudocode}
\usepackage{tikz}
\usepackage{enumitem}   
\usetikzlibrary{fit,positioning,arrows,automata}
\usepackage{setspace}
\usepackage{rotating}
\usepackage{xcolor,colortbl} 
\usepackage{threeparttablex,multirow}
\usepackage{comment}
\usepackage[title]{appendix}
\usepackage[colorlinks=true,linkcolor=blue,urlcolor=blue]{hyperref}
\usepackage{url} % not crucial - just used below for the URL 
\usepackage[colorlinks=true,linkcolor=blue,urlcolor=blue]{hyperref}

\newcommand{\V}[1]{\ensuremath{\boldsymbol{#1}}\xspace}
\newcommand{\ccolor}{\cellcolor[HTML]{C0C0C0}}
\usepackage{authblk}

\newtheorem{thm}{Theorem}
\newtheorem{lemma}{Lemma}

\newtheorem{defn}{Definition}
\def\twoImages#1#2#3#4#5#6 
{
	\centerline{\hfill\makebox[#2]{#3}\hfill\makebox[#5]{#6}\hfill}
	\centerline{\hfill
		\includegraphics[width=#2]{#1}
		\hfill
		\includegraphics[width=#5]{#4}
		\hfill}
}
\definecolor{Gray}{gray}{0.9}

\newenvironment{remark}[1][Remark]{\begin{trivlist}
		\item[\hskip \labelsep {\bfseries #1}]}{\end{trivlist}}

\DeclareMathOperator*{\argmax}{arg\,max}

\newtheorem{manualtheoreminner}{Theorem}
\newenvironment{manualtheorem}[1]{%
	\renewcommand\themanualtheoreminner{#1}%
	\manualtheoreminner
}{\endmanualtheoreminner}

\newtheorem{manuallemmainner}{Lemma}
\newenvironment{manuallemma}[1]{%
	\renewcommand\themanuallemmainner{#1}%
	\manuallemmainner
}{\endmanuallemmainner}

\newcommand{\thistheoremname}{}
\newtheorem*{genericthm*}{\thistheoremname}
\newenvironment{namedthm*}[1]
{\renewcommand{\thistheoremname}{#1}%
	\begin{genericthm*}}
	{\end{genericthm*}}

\usepackage{fancyhdr}
\newcommand\blfootnote[1]{%
  \begingroup
  \renewcommand\thefootnote{}\footnote{#1}%
  \addtocounter{footnote}{-1}%
  \endgroup
}

\begin{document}

%\title{Identifiability and Consistency of Network Inference Using the Hub Model and Variants}
\title{Network Inference Using the Hub Model and Variants \blfootnote{This paper is published in the \emph{Journal of the American Statistical Association (Theory and Methods)}, available on \url{https://doi.org/10.1080/01621459.2023.2183133}.}}

\date{}
%\date{\today}

\author[1]{Zhibing He}
\author[1,*]{Yunpeng Zhao}
\author[2]{Peter Bickel}
\author[3]{Charles Weko}
\author[1]{Dan Cheng}
\author[4]{Jirui Wang}

\affil[1]{Arizona State University} 
\affil[2]{University of California, Berkeley}
\affil[3]{U.S. Army}
\affil[4]{Medpace}
\affil[*]{Corresponding author. Email: Yunpeng.Zhao@asu.edu}

\renewcommand\Authands{ and }

\maketitle

\begin{abstract}
	Statistical network analysis primarily focuses on  inferring  the parameters of an observed network. In many applications, especially in the social sciences, the observed data is the groups  formed by individual subjects.   In these applications, the network is itself a parameter of  a statistical model. \citet{zhao2019network} propose a model-based approach, called the \textit{hub model}, to infer implicit networks from grouping behavior. The  hub model  assumes  that  each member of the  group is brought together by a member of the group called the \textit{hub}. The set of members which can serve as a hub is called the \textit{hub set}. The hub model belongs to the family of Bernoulli mixture models.  Identifiability of Bernoulli mixture model parameters is a notoriously difficult problem. This paper proves  identifiability of the hub model parameters  and estimation consistency under mild conditions. Furthermore,  this paper generalizes the hub model by  introducing a  model component  that allows  hubless groups in which individual nodes  spontaneously appear  independent of any other individual.  We refer to this additional component as the \textit{null component}.  The new model bridges the gap between the hub model and the degenerate case of the mixture model -- the Bernoulli product.  Identifiability and consistency are  also proved for the new model. In addition, a penalized likelihood approach is proposed to estimate the hub set when it is unknown. 
\end{abstract}


\section{INTRODUCTION}\label{sec:intro}

In recent decades, network analysis has been applied in  science and engineering fields including mathematics, physics, biology, computer science, social sciences and statistics (see \cite{Getoor2005, Goldenberg2010, Newman2010} for reviews). Traditionally, statistical network analysis deals with parameter estimation of an observed network, i.e., an observed adjacency matrix.
For example, community detection, a topic of broad interest, studies how to partition the node set of an observed network into cohesive overlapping or non-overlapping communities (see \cite{abbe2017community,zhao2017survey} for recent reviews). Other well-studied statistical network models include the preferential attachment model \citep{Barabasi&Albert1999}, exponential random graph models \citep{Frank&Strauss1986,robins2007introduction}, latent space models \citep{Hoff2002,Hoff2007}, and the graphon model \citep{diaconis2007graph,gao2015rate,zhang2017estimating}.  

In contrast to traditional statistical network analysis, this paper focuses on  inferring a latent network structure. Specifically, we model data with the following format: each observation in the dataset is a subset of nodes that are observed simultaneously.  An observation is called a \textit{group} and  a full dataset  is called \textit{grouped data}.  \cite{Wasserman94} introduced this format using the toy example of a children's birthday party. In  their simple example, children are treated as nodes and  each party represents a group -- i.e., a subset of children who attended the same party is a group. The reader is referred to \cite{zhao2019network,weko2017penalized} for applications of such data to the social sciences and animal behavior. 

The observed grouping behavior presumably results from a latent social structure that can be interpreted as a  network structure of associated individuals \citep{Moreno34}. The  task is therefore to infer  a latent network structure from grouped data. Existing methods  mainly focus on ad-hoc descriptive approaches from the social sciences literature, such as the co-occurrence matrix \citep{Wasserman94} or the half weight index \citep{Cairns87}.  \cite{zhao2019network} propose the first model-based approach, called the \textit{hub model}, which assumes that every observed group has a \textit{hub} that brings together the other members of the group. When the hub nodes of grouped data are known, estimating the model parameters is a trivial task.  In most research situations, hub nodes are unknown and need to be modeled as latent variables. Under this setup, estimating the model parameters becomes a more difficult task.

 %\cite{zhao2019network} demonstrated the hub model by analyzing co-sponsorship of legislation in the Senate of the 110$^{th}$ United States Congress.  The rules of the Senate require that each piece of legislation have a single, unique sponsor; however, other members may co-sponsor the bill.  These rules mean that the legislation sponsorship data conforms to the hub model assumption that every group has a single  hub. Analyzing this data is trivial when the sponsors are known, i.e., when the hubs are observed; \cite{zhao2019network} also estimated the latent network when sponsorship data was eliminated from the data. In general, 

This paper has three aims: first, to prove the identifiability of the canonical parameters and the asymptotic consistency for the estimators of those parameters \textit{when hubs are unobserved}. The canonical parameters refer to the probabilities of being a hub node of a group and the probabilities of being included in a group  formed by a particular hub node. The hub model is a restricted class from the family of finite mixtures of multivariate Bernoulli \citep{zhao2019network}.  \cite{gyllenberg1994non} showed that in general the parameters of finite mixture models of multivariate Bernoulli are  not identifiable.  \cite{zhao2019network} showed that the parameters are identifiable under two assumptions:  the hub node of each group always appears in the group it  forms and relationships are reciprocal. That is, the adjacency matrix is symmetric with diagonal entries as one. This paper considers identifiability when adjacency matrices are asymmetric. The model is therefore referred as to the \textit{asymmetric hub model}.  We prove that when the hub set (i.e., the set of possible hubs) contains at least one fewer member than the node set, the parameters are identifiable under mild conditions. The new setup is practical and less restrictive than the symmetry assumption. Moreover, allowing the hub set to be smaller than the node set can reduce model complexity as pointed out by \cite{weko2017penalized}. When proving the consistency of the estimators, we first prove the consistency of the hub estimates and then show that the estimators of model parameters are consistent as a corollary. Our proofs accommodate the most general setup in which the number of groups (i.e., sample size), the size of the node set, and the size of the hub set are all allowed to grow. 

The second aim is to generalize the hub model  to accommodate  hubless groups and then prove identifiability and consistency of this generalized model. The classical hub model requires each group to have a  hub. As observed in \cite{weko2017penalized}, when fitting the hub model to data, one sometimes has to choose an unnecessarily large hub set due to this requirement. For example, a node that appears infrequently in general but appears once  as a singleton  must be included in the hub set. To relax the \textit{one-hub} restriction, we add a  component to the hub model that allows  hubless groups in which nodes appear independently. We call this additional component the \textit{null component} and call the new model the \textit{hub model with the null component}. %The null component creates a natural connection between the hub model and a null model. That is, if the hub set is empty then the model degenerates to the model in which nodes appear independently in groups. That is, each group is generated by independent Bernoulli trials. 
The proofs of identifiability and consistency for the new model do not parallel the first set of proofs and are more challenging. 

Since the new models assume the hub set is a subset of the nodes, this raises a natural question: how to estimate the hub set from  data, which is the third aim of the paper. We formulate this problem as model selection for Bernoulli mixture models. We borrow the log penalty  in \citet{huang2017model}, originally designed for Gaussian mixture models, to propose a penalized likelihood approach to select the hub set for the hub model with the null component. Instead of penalizing the mixing probability of every component as in \citet{huang2017model}, we modify the penalty function such that the probability of the null component is not penalized. The null component does not exist in the setup of Gaussian mixture models, but it creates a natural connection between the hub model and a null model in our scenario.
That is, when all other mixing probabilities are shrunken to zero, the model naturally degenerates to the model in which nodes appear independently in a group -- in other words, each group is modeled by independent Bernoulli trials.

\section{HUB MODEL AND VARIANTS}
\subsection{Model setup} \label{sec:classical_setup}
First, we review the grouped data structure and then propose a modified version of the hub model, called the \textit{asymmetric hub model}. For a set of $n$ individuals, $V=\{1,\dots, n\}$, we observe $T$ subsets, called \textit{groups}.

In this paper,  groups are treated as a random sample of size $T$ with each group being an  observation. Each  group  is represented by an $n$ length row vector $G^{(t)}$, where
\[ G_i^{(t)} = \left\{ 
\begin{array}{l l}
1 & \quad \textnormal{if node $i$ appears in group $t$,}\\
0 & \quad \textnormal{otherwise, }
\end{array} \right.\]
for $i=1,\dots,n$ and $t=1,\dots,T$. 
The full dataset is a $T \times n$ matrix $\V{G}$ with $G^{(t)}$ being its rows. 

Let $V_0$ be the set of all nodes which can serve as a hub and let $n_L=|V_0|$.  We refer to $V_0$ as the \textit{hub set} and call the nodes in this set \textit{hub set member}. In contrast to  the setup in \cite{zhao2019network} where the  hub set contains all nodes, we assume  that the hub set contains fewer members than  the whole set of nodes, i.e., $n_L<n$. We assume in this section that $V_0$ is known and consider the problem of estimating $V_0$ in Section \ref{sec:unknown_set}.  For  simplicity of notation, we further assume $V_0=\{1,\dots,{n_L}\}$ in this section. We refer to nodes from $n_L+1$ to $n$ as \textit{followers}. Given this notation,  the true hub of $G^{(t)}$ is represented by $z_{*}^{(t)}$ which takes on values from $1,\dots,n_L$.

Under the hub model, each group $G^{(t)}$ is independently generated by the following two-step process:
\begin{enumerate}[label=(\roman*)]
	\item The hub is sampled from a multinomial trial with parameter $\rho=(\rho_1,\dots,\rho_{n_L})$, i.e., $\mathbb{P}(z_*^{(t)}=i)=\rho_i$, with $\sum_{i=1}^{n_L} \rho_i=1$.
	
	\item Given  the hub node $i$,  each node $j$ appears in the group independently with probability $A_{ij}$, i.e., $	\mathbb{P}(G_j^{(t)}=1|z_*^{(t)}=i)=A_{ij}$.  
\end{enumerate}
%Before proceeding, there are a number of implications of the proceeding terms and notation.  We interpret $\rho_i$ to be the probability that node $i$ is the hub of a group and $A_{ij}$ to be the probability that node $j$ is a member of a group given node $i$ is the hub of the group.  Thus, the term \textit{hub set member} applies to any node $i$ with a non-zero $\rho_i$ and the term \textit{follower} applies to any node $j$ with $\rho_j=0$.  Additionally, 
Note that multiple hub set members may appear in the same group although only one of them will be the hub of that group.  %That is, for two nodes $i$ and $j$ where $\rho_i>0$ and $\rho_j>0$, $A_{ij}$ may be non-zero.  

%As in \cite{zhao2019network}, when the hub node of every group is known, estimation of $\{A,\rho\}$ is trivial.  Thus, a key feature of parameter estimation when the hub nodes are unknown is estimating the hub node for each group.  

A key assumption  from \cite{zhao2019network} which we adopt in this paper is that a hub node must appear in any group that it forms (i.e., $A_{ii}\equiv 1$, for $i=1,\dots,n_L$). The parameters for the hub model are thus
\begin{align*}
\rho & =(\rho_1,\dots,\rho_{n_L}), \\
A_{n_L\times n} & = \begin{pmatrix}
1 &  A_{12} & \cdots & A_{1, n_L}  & A_{1,n_L+1} & \cdots & A_{1,n}\\
A_{21} & 1 & \cdots & A_{2,n_L} & A_{2,n_L+1} & \cdots & A_{2,n} \\
\vdots & \vdots & \ddots & \vdots & \vdots & \ddots & \vdots  \\ 
A_{n_L,1} & A_{n_L,2} & \dots & 1 & A_{n_L,n_L+1} & \cdots & A_{n_L,n}
\end{pmatrix}.
\end{align*}
As in \cite{zhao2019network}, we interpret $A_{ij}$ as the strength of the relationship between node $i$ and $j$. We differ from \cite{zhao2019network} in that   $A$ is a non-square matrix and $A_{ij}$ is not necessarily equal to $A_{ji}$. The setting in this article is more natural.   Social relationships are usually non-reciprocal and in most organizations there  are members who do not have the authority or  willingness to initiate  groups. %Later in this paper, we will show how this new setup presents challenges for theoretical analysis but is feasible.

We begin with the case where both $\V{G}$ and $z_*=(z_*^{(1)},\dots,z_*^{(T)})$ are observed. The likelihood function  is
\begin{equation*}
\mathbb{P}(\V{G},z_*|A,\rho)=\prod_{t=1}^T \prod_{i=1}^{n_L} \prod_{j=1}^n \big[A_{ij}^{G_j^{(t)}} (1-A_{ij})^{(1-G_j^{(t)})}\big]^{1(z_*^{(t)}=i)}\prod_{i=1}^{n_L}\rho_i^{1(z_*^{(t)}=i)},
\end{equation*}
where $1(\cdot)$ is the indicator function. With both $\V{G}$ and $z_*$ being observed, it is straightforward to estimate $A$ and $\rho$ by their respective maximum likelihood estimators (MLEs):
\begin{align*}
\hat{A}_{ij}^{z_*}= & \frac{\sum_t G_j^{(t)} 1(z_*^{(t)}=i) }{\sum_t 1(z_*^{(t)}=i)}, \quad  i=1,\dots,n_L,j=1,\dots,n, \\
\hat{\rho}_i^{z_*}= & \frac{\sum_t 1(z_*^{(t)}=i)}{T}, \quad i=1,\dots,n_L.
\end{align*}
%We define $\hat{A}_{ij}^{z_*} = 0$ if $\sum_t 1(z_*^{(t)}=i) = 0$.

When the hub node of each group is latent, i.e., when $z_*$ is unobserved, the estimation problem becomes challenging. Integrating out $z_{*}$, the marginal likelihood of $\V{G}$ is
\begin{equation}\label{likelihood_classical}
\mathbb{P}(\V{G}|A,\rho)=\prod_{t=1}^T \sum_{i=1}^{n_L} \rho_i \prod_{j=1}^n {A_{ij}^{G_j^{(t)}} (1-A_{ij})^{1-G_j^{(t)}}},
\end{equation}
which has the form of a Bernoulli mixture model. Hereafter the term hub model refers to the case where $z_*$ is unobserved, unless otherwise specified. %As will be seen in Section \ref{sec:classical_consistency}, a key step of parameter estimation when the hub nodes are unknown is estimating the hub node for each group. 

%Before considering estimation  of $\rho$ and $A$, we need to  establish the identifiability of parameters $\rho$ and $A$ under \eqref{likelihood_classical}. If there exist two different sets of parameters that can give the same likelihood then the parameters are not estimable.  \cite{zhao2019network} observed that when every node is allowed to be a  hub set member (i.e., $n_L=n$), $\rho$ and $A$ are not identifiable without additional restrictions. They further proved that the symmetry of $A$ is a sufficient condition for identifiability. We remove this stringent constraint and seek a set of milder identifiability conditions in the next section. 

Although less stringent than the original symmetric hub model, the asymmetric hub model has a significant limitation: it cannot naturally transition to a null model. 
In general, a null model generates data that match the basic features of the observed data, but which is otherwise a random process without structured patterns. In other words, a null model is the  degenerate case of the model class being studied. %For example,  a regression line with all regression coefficients being zero except the intercept can be viewed as a null model of multiple linear regression. 
%the Erd\H{o}s-R\'{e}nyi random graph is the null model of stochastic block models (SBMs), i.e,  the SBM with only one community. The Newman--Girvan modularity \citep{Newman&Girvan2004} uses the configuration model as the null model in the criterion function for community detection. In regression analysis,
The null model for grouped data, naturally, generates each group by independent Bernoulli trials. That is, if the grouping behavior is not governed by a network structure then every node is assumed to appear  independently in a group. The likelihood of $G^{(t)}$ under the null model is
\begin{equation*}
\mathbb{P}(G^{(t)})=\prod_{j=1}^n {\pi_{j}^{G_j^{(t)}} (1-\pi_{j})^{1-G_j^{(t)}}},
\end{equation*}
where $\pi_j$ is the probability that node $j$ appears in a group. 

The asymmetric hub model needs generalization to accommodate the null model because if there is only one component in \eqref{likelihood_classical}, say, node $i$ is the only hub set member, the likelihood of $G^{(t)}$ becomes 
\begin{equation*}
\mathbb{P}(G^{(t)})=\prod_{j=1}^n {A_{ij}^{G_j^{(t)}} (1-A_{ij})^{1-G_j^{(t)}}},
\end{equation*}
which is not a proper null model because the assumption $A_{ii} \equiv 1$ forces node $i$ to appear in every group.

To  allow the hub model to degenerate to the null model, we add the null component.  This null component  allows groups without hubs where nodes independently appear in such groups. We call this model the \textit{hub model with the null component}. We use $z_{*}^{(t)}=0$ to represent a hubless group. 
%Under the hub model with the null component, each group $G^{(t)}$ is independently generated by the following two steps:
%\begin{enumerate}[label=(\roman*)]
%	\item The hub is sampled from a multinomial trial with parameter $\rho=(\rho_0,\rho_1,\dots,\rho_{n_L})$, i.e., $\mathbb{P}(z^{(t)}=i)=\rho_i$, with $\sum_{i=0}^{n_L} \rho_i=1$.
	
%	\item If $z^{(t)}=i \in \{1,\dots,n_L\}$, then node $j$ will appear in the group independently with probability $A_{ij}$, i.e., $	\mathbb{P}(G_j^{(t)}=1|z^{(t)}=i)=A_{ij}$. If $z^{(t)}=0$, each node will independently join the group with probability $\pi_j$. 
%\end{enumerate}
%Note that the above model degenerates to the null model when $\rho_0=1$. As before we assume $A_{ii}\equiv 1$ for $i=1,\dots,n_L$.  
The parameters for the hub model with the null component are
$\rho =(\rho_0,\rho_1,\dots,\rho_{n_L}), 
A_{(n_L+1)\times n}=[A_{ij}]_{i=0,1,\dots,n_L,j=1,\dots,n}$.
Here the row indices of $A$ start from 0, i.e., $A_{0j}\equiv \pi_j$ for $j=1,\dots,n$. We will use $A_{0j}$ and $\pi_j$ interchangeably below. As before we assume $A_{ii}\equiv 1$ for $i=1,\dots,n_L$.  
The marginal likelihood of $\V{G}$ under the new model is
\begin{equation}\label{likelihood_null}
\mathbb{P}(\V{G}|A,\rho)=\prod_{t=1}^T \sum_{i=0}^{n_L} \rho_i \prod_{j=1}^n {A_{ij}^{G_j^{(t)}} (1-A_{ij})^{1-G_j^{(t)}}}.
\end{equation}
The above model degenerates to the null model when $\rho_0=1$.
For  simplicity of notation, we use the same notation such as $\rho$ and $A$ for both the hub model with and without the null component when the meaning is clear from context. 

The new model has an advantage in data analysis in addition to  the theoretical benefit. Grouped data usually contain a number of tiny groups such as singletons and doubletons. When fitting the asymmetric hub model to such a dataset, one sometimes has to include these nodes in the hub set due to the one-hub restriction. %For example, as will be shown in Section XX, a singleton must be included in the hub set and at least one node of a doubleton must be included, no matter how infrequently they appear in the dataset, 
Doing so may result in an unnecessarily large hub set (see Section 4 in the Supplemental Materials). In the hub model with the null component, these small groups can be treated as  hubless groups and the corresponding nodes may be removed from the hub set. Therefore, the model complexity is  significantly reduced. 

\subsection{Model identifiability}
Before considering estimation of $\rho$ and $A$ under \eqref{likelihood_classical} and \eqref{likelihood_null}, we need to  establish the identifiability of parameters $\rho$ and $A$. %If there exist two different sets of parameters that can give the same likelihood then the parameters are not estimable.  
\cite{zhao2019network} proved model identifiability under the symmetry condition. We seek a new set of identifiability conditions as the new models do not assume symmetry of $A$.

To precisely define identifiability, let $\mathcal{P}$ be the parameter space of the hub model with the null component, where  $\mathcal{P}=\{(\rho,A) | 0< \rho_i < 1, i=0,\dots,n_L; A_{ii} = 1,i=1,\dots,n_L; 0\leq A_{ij}\leq 1, i=0,\dots,n_L, j=1,\dots,n, i\neq j   \}$. The parameter space of the hub model without the null component is similar except that the index $i$ always begins   with 1.  Let $\V{g}=(g_j^{(t)})_{t=1,\dots,T,j=1,\dots,n}$ be any realization of $\V{G}$ under the hub model.
\begin{defn}\label{def:identi}
	The parameters $(\rho,A)$ within the parameter space $\mathcal{P}$ are identifiable (under the hub model with or without the null component) if the following holds:
	\begin{align*}
	\forall \V{g}, \forall (\tilde{\rho},\tilde{A}) \in \mathbb{P}(\V{G}=\V{g}|\rho,A)= \mathbb{P}(\V{G}=\V{g}|\tilde{\rho},\tilde{A}) \iff  (\rho,A) = (\tilde{\rho},\tilde{A}).
	\end{align*} 
\end{defn} 
We define identifiability in the strictest sense  and the above definition does not allow label swapping of latent classes.  In cluster analysis label swapping refers to the fact that nodes can be successfully partitioned into latent classes, but individual classes cannot be uniquely identified.  For example, community detection may correctly partition voters into communities based on their political preferences, but cannot identify which political party each community prefers.  This is not an issue in the hub model due to the constraint $A_{ii}=1$. In addition, note that we only need to consider identifiability for the distribution of a single observation, i.e., $T=1$ because the data are independently and identically distributed. Let $g$ be a realization of a single observation hereafter.

We now give the identifiability result for the asymmetric hub model. 

\begin{thm}\label{thm:hub_iden}
	The parameters $(\rho,A)$ of the asymmetric hub model are identifiable under the following conditions:
	\begin{enumerate}[label=(\roman*)]
		%\item $0<\rho_i<1$, for $i=1,\dots,n_L$;
		\item $A_{ij}<1$, for $i=1,\dots,n_L, j=1,\dots,n, i\neq j$;
		\item for all $i=1,\dots,n_L$, $i'=1,\dots,n_L,i\neq i'$, the vectors $(A_{i,n_L+1},A_{i,n_L+2},\dots, A_{i,n})$ and $(A_{i',n_L+1},A_{i',n_L+2},\dots, A_{i',n})$ are not identical.
			\end{enumerate}
\end{thm}
Condition (\romannumeral 2) implies that for any pair of  nodes in the hub set, there exists a follower with different probability of being included in groups formed by the two hubs, respectively. All proofs are given in the Supplementary Materials. 

Identifiability under the model with the null component is  more difficult to prove than the case of the asymmetric hub model due to the extra null component in the model. In particular,  there is no constraint such as $\pi_i=1$ on  parameters of the null component. The conditions for identifiability in the following theorem are; however, as natural as those in Theorem \ref{thm:hub_iden}.
\begin{thm}\label{thm:hub_iden_null}
	The parameters $(\rho,A)$ of the hub model with the null component are identifiable under conditions (\romannumeral 1) and (\romannumeral 2) in  Theorem \ref{thm:hub_iden}  (index $i$ begins with 0 in (\romannumeral 1)), and
	\begin{enumerate}[label=(\roman*)]
	  \setcounter{enumi}{2}
		\item for any $i=1,\dots,n_L$, the vectors $(A_{i,n_L+1},A_{i,n_L+2},\dots,A_{i,n})$ 
		and         $(\pi_{n_L+1},\pi_{n_L+2},\dots,\pi_{n})$ are different by at least two entries.

	\end{enumerate}
\end{thm}

Condition (\romannumeral 3) adds the requirement that for any hub $i$, there exist two followers  which each has different probabilities of appearing in a group led by  hub $i$  than of appearing in a  hubless group. This condition implies that there should exist at least two more nodes in the  node set than in the hub set. This condition is  natural if one compares it to condition (\romannumeral 2), as both imply that there exists at least one more column than rows in $A$.

\subsection{Consistency of the maximum profile likelihood estimator}\label{sec:consistency}

We consider the asymptotic consistency for the hub model in the most general setting. That is, we allow the number of groups ($T$), the size of the node set ($n$), and the size of the hub set ($n_L$) to grow. As mentioned in Section \ref{sec:intro}, we  reformulate the problem as a clustering problem where a cluster is defined as the groups  formed by the same hub node.  We borrow the techniques from the community detection literature to prove the consistency of class labels, i.e., the consistency of hub labels. The consistency of parameter estimation then holds as a corollary. Note that $n$ is necessarily to go to infinity for proving the consistency of hub labels because when $n$ is fixed, the posterior probability of the hub label of a group given the data cannot concentrate on a single node. If one is only interested in the consistency of parameter estimation, it is possible to allow $n$ fixed. The problem degenerates to the classical case, that is, estimating a non-growing number of parameters, and the classical theory of MLE is expected to be applicable. 

We first consider the asymmetric hub model without the null component. Let $z=(z^{(t)})_{t=1,\dots,T}$ be an assignment of hub labels. Given $z$, the log-likelihood of the full dataset $\V{G}$ is
\begin{align}\label{cond_lik}
L_G(A|z)=\sum_{t=1}^T \sum_{j=1}^n G_j^{(t)} \log A_{z^{(t)},j}+ (1- G_j^{(t)} ) \log (1-A_{z^{(t)},j} ).
\end{align}
For $i=1,\dots,n_L$, let $t_i=\sum_t 1(z^{(t)}=i)$  be the number of groups with hub $i$. 
Given $z$, the MLE of $A$ is  
\begin{align*}
\hat{A}_{ij}^z=\frac{\sum_t G_j^{(t)} 1(z^{(t)}=i) }{t_i },\,\, \textnormal{for $t_i>0$}.
\end{align*}
If $t_i=0$, define $\hat{A}_{ij}^z=0$.
We will omit the upper index $z$ when it is clear from the context.   Plugging $\hat{A}_{ij}$ back into \eqref{cond_lik}, we obtain the profile log-likelihood 
\begin{align*}
L_G(z)=\max_{A} L_G(A|z)=\sum_t \sum_j G_j^{(t)} \log \hat{A}_{z^{(t)},j}+ (1- G_j^{(t)} ) \log (1-\hat{A}_{z^{(t)},j} ).
\end{align*}
Furthermore, let 
\begin{align*}
\hat{z}=\argmax_{z} L_G(z).
\end{align*}
The framework of profile likelihoods are adopted from the community detection literature \citep{Bickel&Chen2009,Choietal2011}, where $z$ is treated as an unknown parameter and we search for the $z$  that optimizes the profile likelihood. 

Recall that $z_*$ is the true class assignment.    We will treat $z_*$ as a random vector to maintain continuity with the previous sub-section. 
%The same consistency result holds when $z_*$ is treated as fixed.   Let $t_{i*}=\sum_t 1(z_*^{(t)}=i)$, for $i=1,\dots,n_L$. In the following, we will directly assume $\frac{Tc_{\textnormal{min}}}{n_L} \leq t_{i*} \leq \frac{Tc_{\textnormal{max}}}{n_L},i=1,\dots,n_L$, which indeed holds with high probability and can be proved by applying Hoeffding's inequality.

Let $P_j^{(t)}=\mathbb{P}(G^{(t)}_j=1|z_{*}^{(t)})=A_{z_*^{(t)},j}$. Then by replacing $G_j^{(t)}$ by $P_j^{(t)}$, we obtain a ``population version'' of $L_G(z)$: 
\begin{align*}
L_P(z)=\sum_t \sum_j P_j^{(t)} \log \bar{A}_{z^{(t)},j}+ (1- P_j^{(t)} ) \log (1-\bar{A}_{z^{(t)},j} ),
\end{align*}
where 
\begin{align}\label{def_A_bar}
\bar{A}_{ij}=\frac{\sum_t P_j^{(t)} 1(z^{(t)}=i) }{t_i },\,\, \textnormal{for } t_i>0.
\end{align}
Otherwise, define $\bar{A}_{ij}=0$.
Let $T_e=\sum_t 1(z_*^{(t)} \neq \hat{z}^{(t)})$ be the number of groups with  incorrect hub labels. As discussed previously, we do not allow label swapping in the definition of $T_e$.  Our aim is to prove
\begin{align*}
T_e/T=o_p(1), \quad \mbox{as } n_L\rightarrow \infty, n\rightarrow \infty, T\rightarrow \infty.
\end{align*}
We make the following assumptions throughout the proof of consistency under the asymmetric hub model: 
\begin{itemize}
    \item [$H_1$:] $Tc_{\textnormal{min}}/{n_L} \leq t_{i*} \leq Tc_{\textnormal{max}}/{n_L}$ for $i=1,\dots,n_L$,  where $t_{i*}=\sum_t 1(z_*^{(t)}=i)$ and $c_{\textnormal{min}}$ and $c_{\textnormal{max}}$ are constants.
    \item [$H_2$:] $A_{ij}=s_{ij}d$ for $i=1,\dots,n_L,j=1,\dots,n$ and $i\neq j$  where $s_{ij}$ are unknown constants satisfying $0<s_{\textnormal{min}} \leq s_{ij} \leq s_{\textnormal{max}}<\infty$  while $d$ goes to zero as $n$ goes to infinity.
    \item [$H_3$:] There exists a set $V_i \subset \{n_L+1,\dots,n\}$ for $i=1,\dots,n_L$ with\footnote{$|\cdot|$ is the cardinality of a set.} $|V_i| \geq v n/n_L$ such that 
$\tau  = \min_{i,i'=1,\dots,n_L,i\neq i',j\in V_i} \,\, (s_{ij} - s_{i'j})$ is bounded away from 0.
\item  [$H_4$:] $A_{ii'}\leq c_0/n_L $ for $i=1,\dots,n_L$, $i'=1,\dots,n_L$, $i\neq i'$, where $c_0$ is a positive constant.
\end{itemize}
 $H_1$  ensures that no hub set members appear too infrequently. The assumption in fact automatically holds with high probability if $(n_L^2 \log n_L)/T=o(1)$, which can be proved by applying Hoeffding's inequality. Here we directly assume the condition for simplicity.  $H_2$ allows the expected density of $A$ to shrink as $n$ grows, which is a common setup in the community literature.  $H_3$ implies that for every  hub set member there exists a set of nodes  that are more likely  to join  groups initiated by this particular hub set member than others.   The size of this set is influenced by $v$ and the magnitude of this preference is influenced by $d$ (since $A_{ij}=d s_{ij}$).  The decay rates of $d$ and $v$, as well as the growth rates of  $n_L$, $n$ and $T$, will be specified in the following consistency results.
$H_4$ is a technical assumption  that  prevents label swapping from influencing the consistency results. 

% The statement above is well-defined even though we assume $z_*$ is a random vector in this treatment.  We prove the consistency of $\hat{z}$ by a typical proof technique for consistency of M-estimators. That is, the consistency of $\hat{z}$ holds by proving a uniform bound for $|L_G(z)-L_P(z)|$ and proving that $T_e/T$ can be bounded by $L_P(z_*)-L_P(\hat{z})$. 

Now we state a lemma that $T_e/T$ is bounded by $L_P(z_*)-L_P(\hat{z})$. That is, $z_*$ is a \textit{well-separated} point of maximum of  $L_P$. The reader is referred to Section 5.2 in \cite{van2000asymptotic} for the classical case of this concept. 
%As in  community detection, we allow the expected density of the network to shrink as $n$ grows. Specifically, for $i=1,\dots,n_L,j=1,\dots,n$ and $i\neq j$, let $A_{ij}=c_{ij}d$ where $c_1 \leq c_{ij} \leq c_2$  with $ c_1$ and $c_2$ as constants while $d$ goes to zero as $n$ goes to infinity, which controls the overall expected density of the network. 
\begin{lemma}\label{thm:separate}
	Under $H_1$ -- $H_4$, %if $(n_L^2 \log n_L)/T=o(1)$, 
	for some positive constant $\delta$, 
	\begin{align*}
	\mathbb{P} \left (\frac{\delta  n_L }{d v nT}(L_P(z_*)-L_P(\hat{z})) \geq \frac{T_e}{T} \right )\rightarrow 1, \quad \mbox{as } n_L\rightarrow \infty, n\rightarrow \infty, T\rightarrow \infty.
	\end{align*}
\end{lemma}
We consider  the most general setup in which $n_L$, $n$, and $T$ all go to infinity in the main text. For the easier case of $n_L$ being fixed, we give the corresponding results (Theorem $3'$ and $4'$ for the asymmetric hub model and Theorem $5'$ and $6'$  for the hub model with the null component) in the Supplementary Materials. Based on  Lemma \ref{thm:separate}, we establish label consistency: 
\begin{thm}\label{thm:label_consistency}
	Under $H_1$ -- $H_4$, if $n_L^2 \log T/(dTv)=o(1)$, $ (\log d)^2 n_L^2\log n_L/(d n v^2)=o(1)$, and $(\log T)^2 n_L^2\log n_L/ (d n v^2)=o(1)$, then
	\begin{align*}
	T_e/T=o_p(1), \quad \mbox{as } n_L\rightarrow \infty, n\rightarrow \infty, T\rightarrow \infty.
	\end{align*}
\end{thm}

The next result  addresses the consistency for parameter estimation of $A$, which is based upon a faster decay rate of $T_e/T$ than Theorem \ref{thm:label_consistency} (see the proof of Theorem \ref{thm:estimation} in the Supplemental Materials for details). 

\begin{thm}\label{thm:estimation}
	Under $H_1$ -- $H_4$, if 
	$n_L\log n/T = o(1)$, $n_L^3 \log T/(dTv)=o(1)$, $ (\log d)^2 n_L^4\log n_L/(d n v^2)=o(1)$, and $(\log T)^2 n_L^4\log n_L/ (d n v^2)=o(1)$, then
	\begin{align*}
	\max_{i\in \{1,\dots,n_L\}, j \in\{1,\dots,n\}} \left|\hat{A}_{ij}^{\hat{z}}-A_{ij} \right| =o_p(1), \quad \mbox{as } n_L \rightarrow \infty, n \rightarrow \infty, T \rightarrow \infty. 
	\end{align*}
\end{thm}

We now establish the consistency for the hub model with the null component. The proofs are more challenging due to the extra null component. 
We make the following assumptions throughout the proofs, parallel to $H_1$ -- $H_4$:
\begin{itemize}
    \item [$H_1^*$:] $Tc_{\textnormal{min}}/{n_L} \leq t_{i*} \leq Tc_{\textnormal{max}}/{n_L}$ for $i=0,\dots,n_L$,  where $t_{i*}=\sum_t 1(z_*^{(t)}=i)$ and $c_{\textnormal{min}}$ and $c_{\textnormal{max}}$ are constants.
    \item [$H_2^*$:] $A_{ij}=s_{ij}d$ for $i=0,\dots,n_L,j=1,\dots,n$ and $i\neq j$  where $s_{ij}$ are unknown constants satisfying $0<s_{\textnormal{min}} \leq s_{ij} \leq s_{\textnormal{max}}<\infty$  while $d$ goes to zero as $n$ goes to infinity.
    \item [$H_3^*$:] There exists a set $V_i \subset \{n_L+1,\dots,n\}$ for $i=1,\dots,n_L$ with $|V_i| \geq v n/n_L$ such that 
$\tau  = \min_{i=1\dots,n_L,i'=0,\dots,n_L,i\neq i',j\in V_i} \,\, (s_{ij} - s_{i'j})$ is bounded away from 0.
\item  [$H_4^*$:] $A_{ii'}\leq c_0/n_L $ for $i=0,\dots,n_L$, $i'=1,\dots,n_L$, $i\neq i'$,  where $c_0$ is a positive constant.
\end{itemize} 
The main difference between the two sets of assumptions is on the range of the indices. For example, index $i$ is from 0 to $n_L$ in $H_1^*$. In particular, $t_{0*}$ is the true number of hubless groups. Index $i$ starts from $1$ in $H_3^*$ because we only define the set $V_i$ for each hub set member $i$ but not for the hubless case.

We need a result on the separation of $L_P(z_*)$ from $L_P(\hat{z})$ which is similar to Lemma \ref{thm:separate}. However, the technique in the original proof cannot be directly applied to the new model. A key step in the proof of Lemma \ref{thm:separate} relies on the fact that we can obtain a non-zero lower bound for the number of correctly classified groups with node $i$ as the hub node in the asymmetric hub model. Specifically, let $t_{ii'}=\sum_t  1(z_*^{(t)}=i, \hat{z}^{(t)}=i')$ for $i = 0,\dots,n_L$, $i' = 0,\dots,n_L$. Thus, $t_{ii}$ is the number of correctly classified groups  where node $i$ is the hub node. For  the asymmetric hub model, we obtain a lower bound for $t_{ii}/t_{i*}\,\, (i=1,\dots,n_L)$ from the fact that a node cannot be  labeled as the  hub of a particular group if the node does not appear in the group. This is due to the assumption $A_{ii} \equiv 1$ for $i=1,\dots,n_L$. For the hub model with the null component,  the lower bound for $t_{ii}/t_{i*}$ cannot be proved by the same technique  because  all groups can be  classified as  hubless groups without violating the assumption $A_{ii}\equiv 1$. 

We take a different path in the proof to overcome this issue and other technical difficulties due to the null component. We first bound $t_{i0}/t_{i*}$ for $i=1,\dots,n$. \begin{lemma}\label{thm:ti0}
Under $H_1^*$ -- $H_4^*$,  if $n_L^4 \log T/(dTv)=o(1)$, $(\log d)^2 n_L^6\log n_L/ (d n v^2)=o(1)$ and $(\log T)^2 n_L^6\log n_L/(d n v^2)=o(1) $, then for all $\eta>0$,
	\begin{align*}
	\frac{t_{i0}}{t_{i*}}\leq\eta, \quad i=1,\dots,n_L,
	\end{align*}
	with probability approaching 1.
\end{lemma}
Based on the result in Lemma \ref{thm:ti0}, we establish the label consistency for the hub model with the null component.
\begin{thm}\label{thm:label_consistency_null}
	Under the conditions of Lemma \ref{thm:ti0}, 
	\begin{align*}
	\frac{T_e}{T} =o_p(1), \quad \mbox{as } n_L\rightarrow \infty, n\rightarrow \infty, T\rightarrow \infty.
	\end{align*}
\end{thm}

We conclude this section by the result on consistency for parameter estimation of $A$ under the hub model with the null component.
\begin{thm}\label{thm:estimation_null}
	Under $H_1^*$ -- $H_4^*$, if $n_L\log n/T = o(1)$, $n_L^5\log T/(d T v)=o(1)$, $(\log d)^2 n_L^8\log n_L/(d n v^2)=o(1)$ and $(\log T)^2 n_L^8\log n_L/ (d n v^2)=o(1)$, then
	\begin{align*}
	\max_{i\in \{0,\dots,n_L\}, j \in\{1,\dots,n\}} \left|\hat{A}_{ij}^{\hat{z}}-A_{ij} \right| =o_p(1), \quad \mbox{as } n_L \rightarrow \infty, n \rightarrow \infty, T \rightarrow \infty. 
	\end{align*}
\end{thm}

\section{THE HUB MODEL WITH THE NULL COMPONENT AND UNKNOWN HUB SET} \label{sec:unknown_set}
\subsection{Model setup} 

The asymmetric hub model (with or without the null component)  assumes that the hub set is a subset of the nodes. The previous section addressed the estimation problem when the hub set is known, but in practice, the hub set is usually not known a priori. In this section, we study the selection of the hub set under the hub model with the null component. 

 Recall that $V_0$ denotes the hub set with $|V_0|=n_L$. In the following, we no longer assume $V_0=\{1,\dots,n_L\}$ and the goal is to estimate $V_0$.
We begin with a known \textit{potential hub set}, denoted by $\bar{V}_0$, which is subset containing all nodes that can potentially serve as hub set members. One might assume that the ideal $\bar{V}_0$ would be the same as the entire node set $V$; however,  to prove identifiability of parameters when the hub set is unknown (see Theorem $S1$ in the Supplemental Materials), we require the potential hub set $\bar{V}_0$ to be smaller than $V$. In practice, this means we have prior knowledge that certain nodes do not play an important role in group formation and are therefore not included in the hub set. Let $M=|\bar{V}_0|$ with $n_L < M < n$. Without loss of generality, assume $\bar{V}_0 = \{1,\dots,M\}$. 

The data generation mechanism is the same as the hub model with the null component. The parameters  are $\rho =(\rho_0,\rho_1,\dots,\rho_M)$, 
$A_{(M+1)\times n}=[A_{ij}]_{i=0,1,\dots,M,j=1,\dots,n}$.
For $i=1,\dots,M$, $\rho_i=0$ if $i \notin V_0$. The corresponding $\{A_{ij}\}_{j=1,\dots,n}$ therefore do not play a role in the model and will not be estimated. If all $\rho_i=0$, $i=1,\dots,M$, the model degenerates to the null model in which nodes appear independently in all groups. The marginal likelihood of $\bm{G}$ is
\begin{align*}%\label{eq:bmm}
		\mathbb{P}(\bm{G}|{A},{\rho}) 
		& =\prod_{t=1}^{T} \sum_{i=0}^{M}\rho_i \prod_{j=1}^{n}A_{ij}^{G^{(t)}_j}(1-A_{ij})^{1-G^{(t)}_j}.
\end{align*}

\subsection{Penalized likelihood}\label{sec:model_selection}

We propose to maximize the following penalized log-likelihood function to estimate $V_0$: 
\begin{align}
	&  {L}({A},{\rho}) - T\lambda\sum_{i=1}^{M}[\log (\epsilon + \rho_i)-\log \epsilon], \label{eq:lp} \\
	& \textnormal{subject to} \,\, \rho_i \geq 0, \,\, i=0,1,\dots,M, \,\, \sum_{i=0}^M \rho_i=1, \nonumber
\end{align}
where
\begin{equation*}
	{L}({A},{\rho}) = \log \mathbb{P}(\bm{G}|{A},{\rho}) = \sum_{t=1}^T\log \left [\sum_{i=1}^M \rho_i \prod_{j=1}^n A_{ij}^{G_j^{(t)}}(1-A_{ij})^{1-G_j^{(t)}} \right ].
\end{equation*} 
$\lambda$ is the tuning parameter which controls the penalty on the mixing weights. 
$\epsilon$ is a small positive number. We use $\epsilon=10^{-8}$ in all numerical studies. 
  The estimated hub set $V_0$ includes node $i$ $(i=1,\dots,M)$ if and only if $\hat{\rho}_i\neq 0$ in the maximizer of \eqref{eq:lp}.  

The penalty function in \eqref{eq:lp} was inspired by a similar penalty function  proposed by \citet{huang2017model} for selecting the number of components in Gaussian mixture models. However, our penalty function has a subtle but substantial difference: the hub node index $m$ in the penalty function begins with 1 instead of 0  -- that is, we do not penalize  the coefficient of the null component $\rho_0$. The model is therefore penalized toward  the null model, i.e., the independent Bernoulli model, when $\lambda$ is sufficiently large. 
%Due to the existence of the null component, our method  naturally penalizes partial number of components, under the constraint of the sum of all mixing probabilities being 1. \tcr{[I don't understand the previous sentence]} In other words, the hub node index $m$ in the penalty function varies from 1 to $M$, that is, we do not penalize  the coefficient of the null component $\rho_0$. Intuitively, our model is penalized toward  the null model, i.e., the independent Bernoulli model, when $\lambda$ is large enough. %By contrast, it is  unclear that the estimated number of clusters is a monotonic function of $\lambda$ in the setup of \citet{huang2013model} and whether the model is eventually penalized towards the degenerated model for a large $\lambda$. 
The penalty function uses $\log (\epsilon + \rho_i)$ instead of $\log \rho_i$ as in \cite{huang2017model}, because $\log (\epsilon + \rho_i)$ will not go to infinity when $\rho_i$ goes to zero, which makes it possible for $\hat{\rho}_i$ to reach exactly zero.

\cite{gu2019learning} studied model selection under another constrained  class of Bernoulli mixture models -- structured latent attribute models (SLAMs). \cite{gu2019learning} proposed a  penalty function similar to  \cite{huang2017model} but with a hard threshold. \cite{huang2017model} and \cite{gu2019learning} proved the selection consistency under their respective assumed models which we will study for our model in future work. That is, in the context of hub models, whether the selected hub set is identical to the true hub set with high probability when the size of the potential hub set ($M$) diverges.

\subsection{Algorithm}

 We propose a modified expectation-maximization (EM) algorithm for optimizing $\eqref{eq:lp}$. 
 
\textbf{Algorithm 1  (Modified EM)} \\
 Iteratively update $\hat{A}$ and $\hat{z}$ by the following E-step and M-step until convergence. \\
 Define $h_{ti} = \mathbb{P}(z^{(t)} = i | \bm{G},A)$ for $t=1,\dots,T$ and $i=0,\dots,M$. 
\begin{itemize}
 	\item[] \textbf{E-step:} Given $\hat{A}$ and $\hat{\rho}$,  
	\begin{align*}
	\hat{h}_{ti} = \frac{\hat{\rho}_i\mathbb{P}(G^{(t)}|z^{(t)}=i,\hat{A})}{\sum_{i=0}^{M}\hat{\rho}_i\mathbb{P}(G^{(t)}|z^{(t)}=i,\hat{A})}, \quad \textnormal{for}\ i = 0,\dots,M.
	\end{align*}

	\item[] \textbf{M-step:} For $i$ such that $\hat{\rho}_i\neq 0$, given $\hat{h}_{ti}$, 
	\begin{align*}
    \hat{A}_{ij} = \frac{\sum_{t=1}^{T}\hat{h}_{ti}G^{(t)}_j}{\sum_{t=1}^{T}\hat{h}_{ti}}, \,\, \textnormal{for} \ j = 1,\dots,n.
	\end{align*}
	\qquad\qquad Update $\hat{\rho}$ by solving the following optimization problem: 
    \begin{align}
		&\hat{\rho}  = \argmax_{\rho} {L}(\hat{A},{\rho}) -T\lambda\sum_{i=1}^{M}\log(\epsilon+{\rho}_i), \label{eq:hatrho}  \\
		& \text{subject to} \,\, {\rho}_i \geq 0, \,\,  i=0,\dots,M, \,\,\sum_{i=0}^M {\rho}_i = 1.\nonumber
\end{align}
\end{itemize}
The only difference between Algorithm 1 and the standard EM algorithm is the update of $\hat{\rho}$ in the M-step. In the standard EM algorithm for the likelihood without the penalty term, $\hat{\rho}_i$ has a closed-form solution, that is,
$\hat{\rho}_i = \sum_{t=1}^T \hat{h}_{ti}/T,\,\, i=0,\dots,M$.
%The algorithms for the other two hub models are just part of this algorithm, because they do not need to select the hub nodes. Specifically, the only difference is $\hat{\rho}_m  = \argmax_{\rho_m} {L}(\hat{A},{\rho})$.
By contrast, \eqref{eq:hatrho} is a non-linear optimization problem with inequality constraints, which we use a numerical technique -- the augmented Lagrange multiplier \citep{rsolnp2015} method to solve the problem. In addition, since \eqref{eq:lp} is a non-convex optimization problem, we use multiple different initial values (20 random initial values are used in this paper) to help guard against local maxima.

\section{NUMERICAL STUDIES}\label{sec:numerical}

\subsection{Numerical studies when the hub set is known}

In this sub-section, we examine the performance of the estimators for the asymmetric hub model and the hub model with the null component when the hub set is known, under varying $n_L$, $n$ and $T$. Hub set selection will be considered  in the next sub-section. The parameters are estimated by the standard EM algorithm and the estimated hub labels are determined according to the largest posterior probabilities.  

For the asymmetric hub model, let $\rho_i$ be  generated independently from $U(0,1)$ and renormalize $\rho_i$ such that $\sum_{i=1}^{n_L} \rho_i = 1$.
Let the size of the node set, $n$, be 100 or 500. We partition the follower set $\{n_L+1,\dots,n\}$ into $n_L$ non-overlapping sets $V_1,\dots,V_{n_L}$. Each set $V_i$ is the set of followers with a preference for hub set member $i$ over other hub set members. As in Theorem \ref{thm:separate}, we assume different ranges of probabilities of joining a group for   followers that prefer a specific hub set member than for followers which do not prefer that  member. Specifically, for $j \in V_i$, the parameters $A_{ij}$ are generated independently from $U(0.2,0.4)$, and for $j \notin V_i$, the parameters $A_{ij}$ are generated independently from $U(0,0.2)$. 
%For the case of sparse $A$, we introduce a scale factor $\alpha$ to control the density of $A$. Specially, $A_{i,j} \sim U(0.2\alpha,0.4\alpha)$ for $j \in V_i$ and  $A_{i,j} \sim U(0,0.2\alpha)$ for $j \notin V_i$, where $\alpha = 0.1,0.2,\dots,1$ ().
The numerical results for sparser $A$ will be given in Section 4 of the Supplemental Materials.
For clarification, we will not use  prior information about how $A$ was generated  in the estimating procedure. That is, we still treat $A$ as unknown fixed parameters in the estimation. We generate these probabilities from uniform distributions for the sole purpose of adding more variations to the parameter setup. In each setup, we consider four different sample sizes, $T=500,1000,1500$ and 2000, and two different values of the size of hub set, $n_L=10$ and 20.

For the hub model with the null component, let the probability of hubless groups $\rho_0=0.2$, and let $\rho_i$ be generated independently from $U(0,1)$ and renormalize $\rho_i$ such that $\sum_{i=1}^{n_L} \rho_i = 0.8$ for $i=1,\dots,n_L$.
For a hubless group, each node will independently join the group with probability $\pi_j \equiv 0.05$ for $j=1,\dots,n$. The setups on $n_L$, $n$, $\{V_1,\dots,V_{n_L} \}$, $A$, $n_L$ and $T$ are identical to the asymmetric hub model case. 

\begin{table}[!htp]
	\caption{Asymmetric hub model results. Mis-labels: the fraction of groups with incorrect hub labels.  RMSE($\hat{A}_{ij}$): average RMSEs when the hub labels are unknown.  RMSE*: average RMSEs when the hub labels are known.}
	\label{Table:simulation1}
	\begin{center}
		%	\resizebox{4.5in}{!}{
		\begin{tabular}{c c c c | c c c }
			\hline
			
			$n_L=10$	& \multicolumn{3}{c|}{$n=100$} & \multicolumn{3}{c}{$n=500$}\\
			
			\hline
			& Mis-labels & RMSE($\hat{A}_{ij}$) & RMSE*    & Mis-labels & RMSE($\hat{A}_{ij}$)  & RMSE*    \\
			$T=500$  &0.0479 & 0.0501	 & 0.0475 		&0.0011	 &0.0483 &0.0483 \\
			$T=1000$ &0.0335 & 0.0344	 & 0.0332 		&0.0000	 &0.0337 &0.0337 \\
			$T=1500$ &0.0295 & 0.0280	 & 0.0272 	    &0.0000	 &0.0274 &0.0274 \\
			$T=2000$ &0.0262 & 0.0243	 & 0.0236 		&0.0000	 &0.0235 &0.0235 \\
			
			\hline
			
			$n_L=20$	& \multicolumn{3}{c|}{$n=100$} & \multicolumn{3}{c}{$n=500$}\\
			
			\hline
			& Mis-labels & RMSE($\hat{A}_{ij}$) & RMSE*    & Mis-labels & RMSE($\hat{A}_{ij}$)  & RMSE*    \\
			$T=500$  &0.2396 &	0.0791	 & 0.0662		&0.0605	 &0.0686  &0.0673	\\
			$T=1000$ &0.1528 &	0.0548	 & 0.0463		&0.0096	 &0.0466  &0.0463	\\
			$T=1500$ &0.1186 &	0.0433	 & 0.0375	    &0.0029	 &0.0380  &0.0379	\\
			$T=2000$ &0.0998 &	0.0366	 & 0.0325		&0.0013	 &0.0328  &0.0328		\\
			
			\hline
			
		\end{tabular}
		%	}
	\end{center}
	
\end{table}
\begin{table}[!htp]
	\caption{Hub model with null component results. Mis-labels: the fraction of groups with incorrect hub labels.  RMSE($\hat{A}_{ij}$): average RMSEs when the hub labels are unknown.  RMSE*: average RMSEs when the hub labels are known.}
	\label{Table:simulation2}
	\begin{center}
		%	\resizebox{4.5in}{!}{
		\begin{tabular}{c c c c| c c c}
			\hline
			
			$n_L=10$	& \multicolumn{3}{c|}{$n=100$} & \multicolumn{3}{c}{$n=500$}\\
			
			\hline
			& Mis-labels & RMSE($\hat{A}_{ij}$) & RMSE*    & Mis-labels & RMSE($\hat{A}_{ij}$)  & RMSE*    \\
			$T=500$  &0.0842	& 0.0542	 & 0.0511 		&0.0058   & 0.0516	& 0.0516 \\
			$T=1000$ &0.0595	& 0.0376	 & 0.0357 		&0.0006   & 0.0362  & 0.0362  \\
			$T=1500$ &0.0512	& 0.0308	 & 0.0294 	    &0.0001	  & 0.0292	& 0.0292  \\
			$T=2000$ &0.0489	& 0.0264	 & 0.0253 		&0.0001	  & 0.0253	& 0.0253   \\
			
			\hline
			
			$n_L=20$	& \multicolumn{3}{c|}{$n=100$} & \multicolumn{3}{c}{$n=500$}\\
			
			\hline
			& Mis-labels & RMSE($\hat{A}_{ij}$) & RMSE*    & Mis-labels & RMSE($\hat{A}_{ij}$)  & RMSE*    \\
			$T=500$  &0.3206 & 0.0839	 & 0.0734		&0.1146	 & 0.0732 & 0.0719	\\
			$T=1000$ &0.2102 & 0.0607	 & 0.0506		&0.0229	 & 0.0510 & 0.0509	\\
			$T=1500$ &0.1598 & 0.0488	 & 0.0411	    &0.0076	 & 0.0418 & 0.0416	\\
			$T=2000$ &0.1419 & 0.0414	 & 0.0355		&0.0022	 & 0.0359 & 0.0359	\\
			
			\hline
			
		\end{tabular}
		%	}
	\end{center}
	
\end{table}

Table \ref{Table:simulation1} and \ref{Table:simulation2} show the performance of the estimators for the asymmetric hub model and the hub model with the null component, respectively. The first measure of performance we are interested in is the proportion of mislabeled groups, $T_e/T$.  As the proportion of mislabeled groups approaches zero, we expect the parameter estimates to approach the accuracy achievable if the hub nodes are known.  The second measure of performance is the RMSE($\hat{A}_{ij}$).  As a reference point, we also provide the RMSE achieved when we treat the hub nodes as known, RMSE*.  All results are averaged by 1000 replicates. 

From the tables, the estimators for the asymmetric hub model generally outperform those for the hub model with the null component as the latter is a more complex model. The patterns within the two tables are, however, similar. First, the performance becomes better as the sample size $T$ grows, which is in line with common sense in statistics. Second, the performance becomes worse as $n_L$ grows because $n_L$ is the number of components in the mixture model, and thus a larger $n_L$ indicates a more complex model. Third, the effect of $n$ is  more complicated: the RMSE* for the case that hub labels are known slightly increases as $n$ grows because the model contains more parameters. What we are interested in is the case where hub labels are unknown, and this is what our theoretical studies focused on. In this case, the RMSE($\hat{A}_{ij}$) significantly improves as $n$ grows. This is because the clustered pattern becomes clearer as the number of followers increases, which is in line with the label consistency results in Section \ref{sec:consistency}.

\subsection{Numerical results for hub set selection}
We study the performance of hub set selection by the penalized log-likelihood \eqref{eq:lp}, which is optimized by the modified EM algorithm (Algorithm 1). 
We use the same settings as the hub model with the null component in the previous sub-section. The only difference is we need to specify the potential hub set $\bar{V}_0 = \{1,\dots,M\}$: we consider $M=80$ for $n = 100$ and $M=80,200$ and 300 for $n=500$.
In each setup,  AIC and BIC are used to select the tunning parameter, $\lambda$. 
Let $\widehat{V}_0$ be the estimate of $V_0$. The performance of hub set selection is evaluated by the true positive rate (TPR) and the false positive rate (FPR), where
\begin{align*}
\textnormal{TPR} = \frac{\sum_{i=1}^M 1(i \in V_0, i\in \widehat{V}_0)}{n_L}, \,\,\textnormal{FPR} = \frac{\sum_{i=1}^M 1(i \notin V_0, i\in \widehat{V}_0)}{M-n_L}.
\end{align*}

\begin{table}[!htp]
    \footnotesize
	\centering
	\caption{TPR and FPR for hub set selection.} 
	\begin{tabular}{ccccc|cccccc}
		\hline
		\multirow{3}{*}{$n_L$} & \multirow{3}{*}{$T$} & \multirow{3}{*}{Parameter tuning}  & \multicolumn{2}{c}{$n=100$} & \multicolumn{6}{c}{$n=500$}  \\ \cmidrule{4-11}
		& & & \multicolumn{2}{c|}{${M}=80$} & \multicolumn{2}{c}{$M=80$} & \multicolumn{2}{c}{$M=200$} & \multicolumn{2}{c}{${M}=300$}\\
		& & & TPR & FPR & TPR & FPR & TPR & FPR & TPR & FPR \\
		\hline
		\multirow{2}{*}{10}&\multirow{2}{*}{1000}&AIC &0.6438 &0.0719 &0.9460 &0.0128 &0.7338 &0.0081 &0.6986 &0.0128  \\
		& &BIC &0.5787 &0.0283 &0.9381 &0.0127 &0.6831 &0.0042 &0.6472 &0.0081  \\
		\multirow{2}{*}{20}&\multirow{2}{*}{1000}&AIC &0.5140 &0.1410 &0.6972 &0.0249 &0.4831 &0.0229 &0.4780 &0.0370  \\
		& &BIC &0.5100 &0.1350 &0.6859 &0.0239 &0.4494 &0.0132 &0.4673 &0.0318  \\
		\multirow{2}{*}{10}&\multirow{2}{*}{2000}&AIC &0.8613 &0.0187 &0.9909 &0.0010 &0.9130 &0.0018 &0.8585 &0.0015  \\
		& &BIC &0.7675 &0.0043 &0.9883 &0.0005 &0.8956 &0.0007 &0.8400 &0.0004  \\
		\multirow{2}{*}{20}&\multirow{2}{*}{2000}&AIC &0.6560 &0.1050 &0.8551 &0.0074 &0.6770 &0.0155 &0.6250 &0.0140  \\
		& &BIC &0.4438 &0.0344 &0.7884 &0.0034 &0.5848 &0.0058 &0.5519 &0.0056 \\ 
		\hline	
	\end{tabular}\label{tbl:scenario_unif}
\end{table}

Table \ref{tbl:scenario_unif} shows the TPR and FPR for hub set selection under various settings.
The patterns in the table with respect to $n_L, n$ and $T$ are similar to  Table \ref{Table:simulation1} and \ref{Table:simulation2}. That is, the  performance of hub set selection is better for smaller $n_L$, larger $n$, and/or  larger $T$.
Among all  settings, the model with $n_L =10, T = 2000$ and $n = 500$ is the simplest for hub set selection purpose, which has the largest TPR and smallest FPR with $\lambda$ selected by either AIC or BIC.
Furthermore, the selection performance becomes worse as $M$ grows because a larger $M$ corresponds to a larger potential hub set  and hence a larger candidate set of models. 

\section{ANALYSIS OF PASSERINE DATA}

We apply the hub model with the null component to analyze a dataset on grouping behavior of passerines  \citep{shizuka2016measuring}.
The dataset includes 63 color-marked passerines in Australia for daily observations, which are 2 scarlet robins (Petroica boodang), 13 striated thornbills (Acanthiza lineata), 26 buff-rumped thornbills (Acanthiza reguloides), 14 yellow-rumped thornbills (Acanthiza chrysorrhoa), 4 speckled warblers (Chthonicola sagittatus), 2 white-throated treecreepers (Cormobates leucophaea), one white-eared honeyeater (Lichenostomous leucotis), and one unkown bird. 
A group is defined as  individuals observed together in a flock, and in total there are 109 groups, i.e., $T = 109$. Species information is summarized in Table \ref{tbl:data}.

\begin{table}[htp]
    \small
	\centering
	\caption{Summary of passerine species}
	\begin{tabular}{llcc}
		\toprule
		Species & Binomial Nomenclature & Number & Label \\
		\midrule
		scarlet robin & Petroica boodang & 2 & $v_1 - v_2$\\
		striated thornbill & Acanthiza lineata & 13 & $v_3 - v_{15}$\\
		buff-rumped thornbill & Acanthiza reguloides & 26 & $v_{16} - v_{41}$\\
		yellow-rumped thornbill & Acanthiza chrysorrhoa & 14 & $v_{42} - v_{55}$\\
		speckled warbler & Chthonicola sagittatus & 4 & $v_{56} - v_{59}$\\
		white-throated treecreeper & Cormobates leucophaea & 2 & $v_{60} - v_{61}$\\
		white-eared honeyeater & Lichenostomus leucotis & 1 & $v_{62}$\\
		unknown & unknown & 1 & $v_{63}$\\
		\bottomrule
	\end{tabular}\label{tbl:data}
\end{table}

\begin{table}[!htp]
	\centering
	\caption{Estimated hub set for passerine data}
	\begin{tabular}{cccccccccc}
		\toprule
		$\lambda$ & $v_7$ & $v_9$ & $v_{10}$ & $v_{20}$ & $v_{30}$ & $v_{33}$ & $v_{37}$ & $v_{42}$ & $v_{46}$\\     
		\midrule
		0.045 & $\ccolor$ & $\ccolor$ & $\ccolor$ & $\ccolor$ & $\ccolor$ & $\ccolor$ & $\ccolor$ & $\ccolor$ & $\ccolor$ \\
		0.050 &  & $\ccolor$ & $\ccolor$ & $\ccolor$ &  & $\ccolor$ &  & $\ccolor$ & \\
%		\rowcolor{Gray}
		0.055 &  & $\ccolor$ &        &  & $\ccolor$ &  &  & $\ccolor$ &  \\
		0.060 &  & $\ccolor$ &         &  & $\ccolor$    &         &  & $\ccolor$   &   \\
		0.065 &         &         &         &         &         &         &  &         &   \\
		\bottomrule
	\end{tabular}\label{tbl:data_anly}		
\end{table}

In the following analysis, we set the potential hub set $\bar{V}_0$ with $M = 55$ as the collection of birds in the first four species (Table \ref{tbl:data}) and the other eight birds belonging to small-scale species as followers\footnote{Nodes $v_1$ and $v_2$ appear frequently so we include them in the potential hub set.}. 
Table \ref{tbl:data_anly} shows the estimated hub set under various $\lambda$ values where a grey block indicates that a node is included in the hub set. As $\lambda$ increases, nodes are removed gradually from the hub set and at $\lambda=0.065$, the hub model degenerates to the null model where the hub set is empty.
The BIC selects $\lambda=0.055$, where the estimated hub set includes $v_9$, $v_{30}$ and $v_{42}$, each belonging to one of the three large-scale species.

\begin{comment}
Researchers may also be interested in the relationship between the selection proportion and the occurrence frequency of each individual. 
Figure \ref{fig:top5} (a) shows the top 5 components for selection proportion when $\lambda=0.025$, which are $v_{26}$, $v_9$, $v_4$, $v_{33}$ and $v_{37}$ in order.
Figure \ref{fig:top5} (b) shows the top 5 components for occurrence frequency, which are $v_2$, $v_1$, $v_{16}$, $v_{42}$ and $v_{26}$ in order.
From the figures, there is no particular relationship between the two measures, which indicates the subtlety of our component selection method, that is, the importance of an individual depends on the grouping structure of all individuals, not only on itself occurrence frequency.

\begin{figure}[!htp]
\centering
\subfloat[]{\includegraphics[width=0.80\textwidth]{fig//Top5_prop.png}}\hspace{-1.5em}%
\subfloat[]{\includegraphics[width=0.80\textwidth]{fig//Top5_freq.png}}
\caption{Top 5 components with respect to selection proportion and occurrence frequency ($\lambda = 0.025$)}
\label{fig:top5}
\end{figure}
\end{comment}

\section{SUMMARY AND DISCUSSION} \label{sec:summary}
In this paper we studied the theoretical properties of the hub model and its variants from the perspective of Bernoulli mixture models. 
The contributions of the paper are four-fold. First, we proved the model identifiability  of the hub model. Bernoulli mixture models are a notoriously difficult model to prove identifiability on, especially under mild conditions. Second, we proved the label consistency and estimation consistency of the hub model. Third, we generalized the hub model by adding the null component that allows nodes to independently appear in hubless groups. The new model can naturally degenerate to the null model -- the Bernoulli product. We also proved identifiability and consistency of the newly proposed model. 
Finally, we  proposed a penalized likelihood method to select the hub set, which estimates not only the size of the hub set, $n_L$, but also which nodes belong to the set. The new method can handle data with no prior knowledge of the hub set and hence greatly expands the domain of the applicability of the hub model.

A natural constraint from \cite{zhao2019network} that we  apply in this paper is $A_{ii}=1 \,\, (i=1,\dots,n_L)$, which turns out to be a key condition for ensuring model identifiability and  avoiding the label swapping issue in the proof of consistency. On the other hand, this constraint prevents the asymmetric hub model from naturally degenerating to the null model because one node always appear in every group when there is only one component in the hub model, which motivated adding the null component to the model. 

 We consider the profile likelihood estimator in the proofs of consistency. The marginal likelihood MLE could also be studied using a different framework. \cite{bickel2013asymptotic} and \cite{brault2020consistency} proved the consistency of the marginal likelihood MLE under the block models for undirected and directed networks, respectively. Their approach is to first prove the consistency of the MLE  under the complete data likelihood and to further show that the marginal likelihood is asymptotically equivalent to the complete data likelihood, which implies the consistency of the MLE under the marginal likelihood. We plan to extend the above framework to the hub model for future works. Moreover, we plan to study the model selection consistency of the proposed hub set selection method, especially when $n_L$, $n$ and $T$ are all allowed to grow.
What we would also like to explore is to go beyond the independence assumption and to develop theories and model selection methodologies for correlated or temporally dependent groups \citep{zhao_temporal}. 

Finally, we briefly review  other work on Bernoulli mixture models. \cite{gyllenberg1994non} first showed that finite mixtures of Bernoulli products are not identifiable. \cite{allman2009identifiability} introduced and studied the concept of generic identifiability, which means that the set of non-identifiable parameters has Lebesgue measure zero. Identifiability under another class of mixture Bernoulli models has been recently studied \citep{xu2017identifiability,gu2019learning,gu2019sufficient}. This class of models, for example, structured latent attribute models (SLAMs), has applications in psychological and educational research. The motivation, the model setup, and the proof techniques presented in this paper  are all different from previous research, and the result of neither implies the other.\cite{gu2019learning} further established the selection consistency in SLAMs when the number of potential latent patterns goes to infinity. It is intriguing to combine the techniques in the present paper and in \cite{gu2019learning} to study the selection consistency in the hub model with the null component, especially for the case that both the size of the true hub set ($n_L$) and the size of the potential hub set ($M$) go to infinity.  
\section*{Supplementary Materials}
The supplementary materials contain proofs of all technical results in the paper, additional numerical studies, and an analysis of extended bakery data.

\section*{Acknowledgements}
The authors thank the editor, the associate editor, and two anonymous referees for their constructive feedback and suggestions.
\section*{Funding}
Yunpeng Zhao acknowledges support from National Science Foundation grant DMS-1840203. Peter Bickel acknowledges support from National Science Foundation grant DMS-1713083. Dan Cheng and Zhibing He acknowledge support from National Science Foundation grant DMS-1902432 Simons Foundation Collaboration Grant 854127.

\clearpage
\appendix   %% just one so don't need appendices environment
\renewcommand{\appendixpagename}{\centering\textbf{\Large Supplementary Materials}}

%\appendixpage %% in article class prodes a \part title (left justified)
\begin{center}  %% center APPENDIX
\appendixpagename
\end{center}

\section{Proofs in Section 2.2}

%%%%%%%%%%%%%%%%%%%%%%%%%------------------Theorem 1
\begin{proof}[Proof of Theorem 1]
	Let $(\tilde{\rho},\tilde{A})\in \mathcal{P}$ be a set of parameters such that $\mathbb{P}(g|\rho,A)= \mathbb{P}(g|\tilde{\rho},\tilde{A})$ for all $g$. 	For all $i=1,\dots,n_L$, $k=n_L+1,\dots,n$, consider the probability that only $i$ appears under parameterizations $(\rho,A)$ and $(\tilde{\rho},\tilde{A})$, respectively
	\begin{align*}
	\tilde{\rho}_i (1-\tilde{A}_{ik}) \prod_{j=1,\dots,n, j \neq i,j \neq k} (1-\tilde{A}_{ij}) & = \rho_i (1-A_{ik}) \prod_{j=1,\dots,n, j \neq i,j \neq k} (1-A_{ij}),
	\end{align*}
	and the probability that only $i$ and $k$ appear
	\begin{align*}
	\tilde{\rho}_i \tilde{A}_{ik}  \prod_{j=1,\dots,n,j \neq i, j \neq k} (1-\tilde{A}_{ij})  & =\rho_i A_{ik} \prod_{j=1,\dots,n,j \neq i, j \neq k} (1-A_{ij}).
	\end{align*}
	As $A_{ij}<1$ in condition (\romannumeral 1), dividing the second equation by the first, we obtain $\tilde{A}_{ik}/(1-\tilde{A}_{ik})=A_{ik}/(1-A_{ik})$ and hence $\tilde{A}_{ik}=A_{ik}$ for $i=1,\dots,n_L$, $k=n_L+1,\dots,n$.

	For any $i=1,\dots,n_L$, $i'=1,\dots,n_L$, $i\neq i'$, suppose that $k$ is the follower such that $A_{ik} \neq A_{i'k}$. Consider the probability that only $i$ and $i'$ appear  
	\begin{align*}
	&\tilde{\rho}_i \tilde{A}_{ii'} (1-\tilde{A}_{ik}) \prod_{j=1,\dots,n,j \neq i, j \neq i',j\neq k}(1-\tilde{A}_{ij})+\tilde{\rho}_{i'} \tilde{A}_{i'i}(1-\tilde{A}_{i'k})\prod_{j=1,\dots,n,j \neq i, j \neq i',j\neq k} (1-\tilde{A}_{i'j})  \\
	= & \rho_i A_{ii'} (1-A_{ik}) \prod_{j=1,\dots,n,j \neq i, j \neq i',j\neq k}(1-A_{ij})+\rho_{i'} A_{i'i}(1-A_{i'k})\prod_{j=1,\dots,n,j \neq i, j \neq i',j\neq k} (1-A_{i'j}),  
	\end{align*}
	and the probability that $i$, $i'$ and $k$ appear
	\begin{align*}
	&\tilde{\rho}_i \tilde{A}_{ii'} \tilde{A}_{ik}\prod_{j=1,\dots,n,j \neq i, j \neq i',j\neq k}(1-\tilde{A}_{ij})+\tilde{\rho}_{i'} \tilde{A}_{i'i}\tilde{A}_{i'k}\prod_{j=1,\dots,n,j \neq i, j \neq i',j\neq k} (1-\tilde{A}_{i'j})  \\
	= & \rho_i A_{ii'} A_{ik} \prod_{j=1,\dots,n,j \neq i, j \neq i',j\neq k}(1-A_{ij})+\rho_{i'} A_{i'i}A_{i'k}\prod_{j=1,\dots,n,j \neq i, j \neq i',j\neq k} (1-A_{i'j}).
	\end{align*}
	As $\tilde{A}_{ik}=A_{ik}$ for $i=1,\dots,n_L$, $k=n_L+1,\dots,n$, the above two equations become
	\begin{align}
	&\tilde{\rho}_i \tilde{A}_{ii'} (1-A_{ik}) \prod_{j=1,\dots,n,j \neq i, j \neq i',j\neq k}(1-\tilde{A}_{ij})+\tilde{\rho}_{i'} \tilde{A}_{i'i}(1-A_{i'k})\prod_{j=1,\dots,n,j \neq i, j \neq i',j\neq k} (1-\tilde{A}_{i'j})  \nonumber \\
	= & \rho_i A_{ii'} (1-A_{ik}) \prod_{j=1,\dots,n,j \neq i, j \neq i',j\neq k}(1-A_{ij})+\rho_{i'} A_{i'i}(1-A_{i'k})\prod_{j=1,\dots,n,j \neq i, j \neq i',j\neq k} (1-A_{i'j}),  \label{t1} \\
	&\tilde{\rho}_i \tilde{A}_{ii'} A_{ik} \prod_{j=1,\dots,n,j \neq i, j \neq i',j\neq k}(1-\tilde{A}_{ij})+\tilde{\rho}_{i'} \tilde{A}_{i'i}A_{i'k}\prod_{j=1,\dots,n,j \neq i, j \neq i',j\neq k} (1-\tilde{A}_{i'j})  \nonumber \\
	= & \rho_i A_{ii'} A_{ik} \prod_{j=1,\dots,n,j \neq i, j \neq i',j\neq k}(1-A_{ij})+\rho_{i'} A_{i'i}A_{i'k}\prod_{j=1,\dots,n,j \neq i, j \neq i',j\neq k} (1-A_{i'j}).   \label{t2} 
	\end{align}
	\eqref{t1} and \eqref{t2} can be viewed as a system of linear equations with unknown variables
	\begin{align*}
	\tilde{\rho}_i \tilde{A}_{ii'}  \prod_{j=1,\dots,n,j \neq i, j \neq i',j\neq k}(1-\tilde{A}_{ij}),
	\end{align*}
	and 
	\begin{align*}
	\tilde{\rho}_{i'} \tilde{A}_{i'i}\prod_{j=1,\dots,n,j \neq i, j \neq i',j\neq k} (1-\tilde{A}_{i'j}).
	\end{align*} 
	By condition (\romannumeral 2), as $A_{ik} \neq A_{i'k}$, the system has full rank and hence has one and only one solution: 
	\begin{align} 
	\tilde{\rho}_i \tilde{A}_{ii'}  \prod_{j=1,\dots,n,j \neq i, j \neq i',j\neq k}(1-\tilde{A}_{ij}) & =\rho_i A_{ii'}  \prod_{j=1,\dots,n,j \neq i, j \neq i',j\neq k}(1-A_{ij}), \nonumber \\
	\tilde{\rho}_{i'} \tilde{A}_{i'i}\prod_{j=1,\dots,n,j \neq i, j \neq i',j\neq k} (1-\tilde{A}_{ij}) & =\rho_{i'} A_{i'i}\prod_{j=1,\dots,n,j \neq i, j \neq i',j\neq k} (1-A_{i'j}).  \label{hub_final}
	\end{align}
	
	Combining \eqref{hub_final} with 
	\begin{align*}
	\tilde{\rho}_i (1-\tilde{A}_{ii'})  \prod_{j=1,\dots,n,j \neq i, j \neq i',j\neq k}(1-\tilde{A}_{ij}) & =\rho_i (1-A_{ii'})  \prod_{j=1,\dots,n,j \neq i, j \neq i',j\neq k}(1-A_{ij}),
	\end{align*}
	we obtain $\tilde{A}_{ii'}=A_{ii'}$ for $i=1,\dots,n_L, i'=1,\dots,n_L$ by a similar argument to that at the beginning of the proof. It follows immediately that $\tilde{\rho}_i=\rho_i$ for $i=1,\dots,n_L$.
\end{proof}

\begin{remark}
Neither conditions in Theorem 1 can be removed. That is, if either condition is removed, then there exists $(\rho,A) \in \mathcal{P}$ such that $(\rho,A)$ is not identifiable. In fact,
\begin{align*}
    \rho=(1/2,1/2), \,\, A=\begin{pmatrix}
    1 & 1/2 & 0 \\
    1 & 1 & 1/2 
    \end{pmatrix}
\end{align*}
and 
\begin{align*}
    \rho=(1/4,3/4), \,\, A=\begin{pmatrix}
    1 & 0 & 0 \\
    1 & 1 & 1/3 
    \end{pmatrix}
\end{align*}
give the same probability distribution, which implies condition (\romannumeral 1) is necessary.

Moreover,
\begin{align*}
        \rho=(1/2,1/2), \,\, A=\begin{pmatrix}
    1 & 1/2 & 1/2 \\
    1/2 & 1 & 1/2 
    \end{pmatrix}
    \end{align*}
    and 
\begin{align*}
    \rho=(1/4,3/4), \,\, A=\begin{pmatrix}
    1 & 0 & 1/2\\
    2/3 & 1 & 1/2
    \end{pmatrix}
\end{align*}
give the same probability distribution, which implies condition (\romannumeral 2) is necessary. 
\end{remark}

%%%%%%%%%%%%%%%%%%%%%%%%%%%%%%%-----------------------Theorem 2
\begin{proof}[Proof of Theorem 2]
	Let $(\tilde{\rho},\tilde{A})\in \mathcal{P}$ be a set of parameters of the hub model with the null component such that $\mathbb{P}(g|\rho,A)= \mathbb{P}(g|\tilde{\rho},\tilde{A})$ for all $g$. 
	Consider the probability that no one appears:
	\begin{align*}
	\tilde{\rho}_0 \prod_{j=1}^n (1-\tilde{\pi}_j) & = \rho_0 \prod_{j=1}^n (1-\pi_j).
	\end{align*}
	For $k=n_L+1,\dots,n$, consider the probability that only $k$ appears:
	\begin{align*}
	\tilde{\rho}_0 \tilde{\pi}_k \prod_{j=1,\dots,n,j\neq k} (1-\tilde{\pi}_j) & =\rho_0 \pi_k \prod_{j=1,\dots,n,j\neq k} (1-\pi_j).
	\end{align*}
	From the above equations, we obtain
	\begin{align} \label{important0}
	& \tilde{\pi}_k  = \pi_k, \quad k=n_L+1,\dots,n \nonumber, \\
	& \tilde{\rho}_0 \prod_{j=1}^{n_L} (1-\tilde{\pi}_j) = \rho_0 \prod_{j=1}^{n_L} (1-\pi_j) .
	\end{align}
	
	By condition (\romannumeral 3), for $i=1,\dots,n_L$, let $k$ and $k'$ be the nodes from $\{n_L+1,\dots,n\}$ such that $\pi_k \neq A_{ik}$ and $\pi_{k'} \neq A_{ik'}$. 
	
	Consider the  probability that $i$ appears but no other nodes from $\{1,\dots,n_L\}$ appears (the rest do not matter)
	\begin{align}
	& \tilde{\rho}_0 \tilde{\pi}_i  \prod_{j=1,\dots,n_L, j \neq i} (1-\tilde{\pi}_j)  + \tilde{\rho}_i \prod_{j=1,\dots,n_L, j \neq i} (1-\tilde{A}_{ij}) \nonumber   \\
	= &  \rho_0 \pi_i  \prod_{j=1,\dots,n_L, j \neq i} (1-\pi_j)  + \rho_i \prod_{j=1,\dots,n_L, j \neq i} (1-A_{ij});  \label{temp1}
	\end{align}
	the probability that $i$ and $k$ appear but no other nodes from $\{1,\dots,n_L\}$ appears (the rest do not matter)
	\begin{align}
	& \tilde{\rho}_0 \tilde{\pi}_i  \prod_{j=1,\dots,n_L, j \neq i} (1-\tilde{\pi}_j) \pi_{k}+ \tilde{\rho}_i \prod_{j=1,\dots,n_L, j \neq i} (1-\tilde{A}_{ij}) \tilde{A}_{ik} \nonumber \\
	= &   \rho_0 \pi_i  \prod_{j=1,\dots,n_L, j \neq i} (1-\pi_j) \pi_{k}+ \rho_i \prod_{j=1,\dots,n_L, j \neq i} (1-A_{ij}) A_{ik}; \label{temp2} 
	\end{align}
	the probability that $i$ and $k'$ appear but no other nodes from $\{1,\dots,n_L\}$ appears (the rest do not matter)
	\begin{align}
	& \tilde{\rho}_0 \tilde{\pi}_i  \prod_{j=1,\dots,n_L, j \neq i} (1-\tilde{\pi}_j) \pi_{k'}+ \tilde{\rho}_i \prod_{j=1,\dots,n_L, j \neq i} (1-\tilde{A}_{ij}) \tilde{A}_{ik'} \nonumber \\
	= &   \rho_0 \pi_i  \prod_{j=1,\dots,n_L, j \neq i} (1-\pi_j) \pi_{k'}+ \rho_i \prod_{j=1,\dots,n_L, j \neq i} (1-A_{ij}) A_{ik'}; \label{temp3} 
	\end{align}
	and the probability that $i,k$ and $k'$ appear but no other nodes from $\{1,\dots,n_L\}$ appears (the rest do not matter)
	\begin{align}
	& \tilde{\rho}_0 \tilde{\pi}_i  \prod_{j=1,\dots,n_L, j\neq i} (1-\tilde{\pi}_j)\pi_k \pi_{k'} + \tilde{\rho}_i \prod_{l=1,\dots,n_L, j \neq i} (1-\tilde{A}_{ij}) \tilde{A}_{ik} \tilde{A}_{ik'}  \nonumber \\
	= &  \rho_0 \pi_i  \prod_{j=1,\dots,n_L, j\neq i} (1-\pi_j)\pi_k \pi_{k'} + \rho_i \prod_{l=1,\dots,n_L, j \neq i} (1-A_{ij}) A_{ik} A_{ik'}.  \label{temp4}
	\end{align}
	Note that the above equations  are not probabilities of a single realization $g$ but are sums of multiple $\mathbb{P}(g)$. Moreover, we put $\pi_k,\pi_{k'}$ instead of $\tilde{\pi}_k,\tilde{\pi}_{k'}$ on the LHS of the equations, since we have proved $\tilde{\pi}_k=\pi_k,k=n_L+1,\dots,n$. 
	
	Let 
	\begin{align*}
	x & = \rho_0 \pi_i  \prod_{j=1,\dots,n_L, j \neq i} (1-\pi_j), \\
	\tilde{x} & =\tilde{\rho}_0 \tilde{\pi}_i  \prod_{j=1,\dots,n_L, j \neq i} (1-\tilde{\pi}_j), \\
	y & = \rho_i \prod_{j=1,\dots,n_L, j \neq i} (1-A_{ij}), \\
	\tilde{y} & =  \tilde{\rho}_i \prod_{l=1,\dots,n_L, j \neq i} (1-\tilde{A}_{ij}). 
	\end{align*}
	Then \eqref{temp1}, \eqref{temp2} \eqref{temp3} and \eqref{temp4} become
	\begin{align*}
	\tilde{x}+\tilde{y} & =x+y,  \\
	\tilde{x}\pi_k +\tilde{y} \tilde{A}_{ik} & = x\pi_k +y  A_{ik},   \\
	\tilde{x}\pi_{k'} +\tilde{y} \tilde{A}_{ik'} & = x\pi_{k'} +y  A_{ik'}, \\
	\tilde{x}\pi_k\pi_{k'}  +\tilde{y} \tilde{A}_{ik} \tilde{A}_{ik'} & = x\pi_k\pi_{k'} +y A_{ik} A_{ik'}. 
	\end{align*}
	
	Plugging $\tilde{x}-x=y-\tilde{y}$ into the last three equations, we obtain 
	\begin{align}
	\label{important1} \tilde{y} \tilde{A}_{ik} & =  \tilde{y} \pi_k+ y (A_{ik}-\pi_k), \\
	\label{important2} \tilde{y} \tilde{A}_{ik'} & =  \tilde{y} \pi_{k'}+ y (A_{ik'}-\pi_{k'}), \\
	\label{important3} y\pi_k\pi_{k'}  +\tilde{y} \tilde{A}_{ik} \tilde{A}_{ik'} & = \tilde{y} \pi_k\pi_{k'} +y A_{ik}  A_{ik'}.
	\end{align}
	Multiplying \eqref{important3} by $\tilde{y}$, and plugging the right hand sides of \eqref{important1} and \eqref{important2} into the resulting equation, we obtain
	\begin{align*}
	& y\tilde{y}\pi_k\pi_{k'}  + \tilde{y}^2 \pi_k \pi_{k'}+\tilde{y} \pi_k y (A_{ik'}-\pi_{k'})+ \tilde{y} \pi_{k'} y (A_{ik}-\pi_k)+y^2 (A_{ik}-\pi_k)(A_{ik'}-\pi_{k'}) \\
	& = \tilde{y}^2 \pi_k\pi_{k'} +y\tilde{y} A_{ik}  A_{ik'}, \\
	\Rightarrow &  y (A_{ik}-\pi_k)(A_{ik'}-\pi_{k'}) = \tilde{y} (A_{ik}-\pi_k)(A_{ik'}-\pi_{k'}).
	\end{align*}
	Therefore,  
	$
	\tilde{y}=y
	$ since $\pi_k \neq A_{ik}$ and $\pi_{k'} \neq A_{ik'}$. It follows that $\tilde{x}=x$, i.e., 
	\begin{align*}
	\tilde{\rho}_0 \tilde{\pi}_i  \prod_{j=1,\dots,n_L, j \neq i} (1-\tilde{\pi}_j) = \rho_0 \pi_i  \prod_{j=1,\dots,n_L, j \neq i} (1-\pi_j), \quad i=1,\dots,n_L.
	\end{align*}
	
	Combining the above equation with \eqref{important0}, we obtain 
	\begin{align*}
	&\tilde{\pi}_i = \pi_i, \quad i=1,\dots,n_L, \\
	& \tilde{\rho}_0=\rho_0.
	\end{align*}

	Note that $\mathbb{P}(g)=\mathbb{P}(g|z=0)\mathbb{P}(z=0)+\mathbb{P}(g|z\neq 0)\mathbb{P}(z\neq 0)$. So far we have proved parameters of $\mathbb{P}(g|z=0)$ and $\mathbb{P}(z=0)$ are identifiable. We only need to prove the identifiability of $\mathbb{P}(g|z\neq 0)$, which is the case of the asymmetric hub model and has been proved by Theorem 1.
\end{proof}
\begin{remark}
No conditions in Theorem 2 can be removed. Here we only give a counterexample when condition (\romannumeral 3) is not satisfied  since the other two are similar to the case of Theorem 1. In fact,
\begin{align*}
        \rho=(1/2,1/2), \,\, A=\begin{pmatrix}
    1/2 & 0 & 1/2 \\
    1 & 1/2 & 1/2 
    \end{pmatrix}
    \end{align*}
    and 
\begin{align*}
    \rho=(1/4,3/4), \,\, A=\begin{pmatrix}
    0 & 0 & 1/2\\
    1 & 1/3 & 1/2
    \end{pmatrix}
\end{align*}
give the same probability distribution. 

\end{remark}
\section{Proofs in Section 2.3}

We start by recalling notations defined in the main text. Recall that $z_{*}$ is the true label assignment, $z$ is an arbitrary label assignment, and $\hat{z}$ is the maximum profile likelihood estimator. Furthermore, $t_{i*}=\sum_t 1(z_*^{(t)}=i)$, and $ t_{i}=\sum_t 1(z^{(t)}=i),t_{ii'}=\sum_t  1(z_*^{(t)}=i, \hat{z}^{(t)}=i' )$.

%%%%%%%%%%%%%%%%%%%%%%%%----------------------Lemma 1

\begin{proof}[Proof of Lemma 1]

We first prove a fact: under $H_1$ and $H_4$, for $0<\delta_1<e^{-c_0}$, 
	\begin{align*}
	\mathbb{P} \left ( \bigcup_{i=1}^{n_L} \left  \{\frac{t_{ii}}{t_{i*}}\leq \delta_1 \right \} \right ) \rightarrow 0.
	\end{align*}
	Note that $\hat{z}$ must be feasible (the estimated hub must appear in the group as we assume $A_{ii}\equiv 1$), we have 
	\begin{align}
	& \mathbb{P} \left ( \left . \frac{t_{ii}}{t_{i*}}\leq \delta_1 \right | z_* \right ) \nonumber \\
	\leq & \mathbb{P} \left ( \left . \frac{1}{t_{i*}}  \sum_{t=1}^T 1(z_*^{(t)}=i)  \prod_{k \in \{1,\dots,n_L \}, k\neq i} (1-G_k^{(t)}) \leq \delta_1 \right | z_* \right ) \label{bern2}.
	\end{align}
Now since 
	\begin{align*}
	\mathbb{E} \left [ \left .  \prod_{k \in \{1,\dots,n_L \}, k\neq i}  (1-G_k^{(t)}) \right | z_*^{(t)}=i \right ] = \prod_{k \in \{1,\dots,n_L \}, k\neq i} (1-A_{ik})   \geq (1-c_0/n_L)^{n_L} \geq e^{-c_0},
	\end{align*}
	by Hoeffding's inequality,
	\begin{align*}
	\eqref{bern2}  \leq & \mathbb{P} \left ( \left . \frac{1}{t_{i*}}  \sum_{t=1}^T 1(z_*^{(t)}=i) \left [ \prod_{k \in \{1,\dots,n_L \}, k\neq i} (1-G_k^{(t)}) -\prod_{k \in \{1,\dots,n_L \}, k\neq i} (1-A_{ik})\right ] \leq \delta_1-e^{-c_0} \right | z_* \right ) \\
	\leq & \exp \{ -2t_{i*} (e^{-c_0}-\delta_1 )^2 \}.
	\end{align*}
	Hence
	\begin{align*}
	& \mathbb{P} \left ( \left .  \bigcup_{i=1}^{n_L} \left \{\frac{t_{ii}}{t_{i*}}\leq \delta_1 \right \} \right | z_* \right ) \\
	= & \mathbb{P} \left ( \left .   \bigcup_{i=1}^{n_L} \left \{\frac{t_{ii}}{t_{i*}}\leq \delta_1 \right \}, \{ t_{i*} \geq c_\textnormal{min} T/n_L, \textnormal{for all $i$} \} \right | z_* \right ) \\
	& + \mathbb{P} \left ( \left .   \bigcup_{i=1}^{n_L} \left \{\frac{t_{ii}}{t_{i*}}\leq \delta_1 \right \}, \{ t_{i*} < c_\textnormal{min} T/n_L, \textnormal{for some $i$} \} \right | z_* \right ) \\
	\leq  &  \sum_{i=1}^{n_L} \mathbb{P} \left ( \left . \frac{t_{ii}}{t_{i*}}\leq \delta_1 \right | z_* \right )1(t_{i*}\geq c_\textnormal{min} T/n_L) \\
	& +1(t_{i*} < c_\textnormal{min} T/n_L, \textnormal{for some $i$}) \\
	\leq &  n_L \exp \{ -2c_\textnormal{min} T/(n_L) (e^{-c_0}-\delta_1 )^2 \} +1(t_{i*} < c_\textnormal{min} T/n_L, \textnormal{for some $i$}).
	\end{align*}
	It follows that
	\begin{align*}
	& \mathbb{P} \left ( \bigcup_{i=1}^{n_L} \left \{\frac{t_{ii}}{t_{i*}}\leq \delta_1 \right \} \right ) \\
	= & \mathbb{E}_{z_*} \left [\mathbb{P} \left ( \left .  \bigcup_{i=1}^{n_L} \left \{\frac{t_{ii}}{t_{i*}}\leq \delta_1 \right \} \right | z_* \right )\right ] \\
	\leq & n_L \exp \{ -2c_\textnormal{min} T/(n_L) (e^{-c_0}-\delta_1 )^2 \} +\mathbb{P}(t_{i*} < c_\textnormal{min} T/n_L, \textnormal{for some $i$}) \rightarrow 0.
	\end{align*}
	Therefore, $\frac{t_{ii}}{t_{i*}}\geq \delta_1 $ for $i=1,\dots,n_L$ with probability approaching 1.
	
	Let $\mathcal{E}=\{ \frac{t_{ii}}{t_{i*}}\geq \delta_1 \textnormal{ and } t_{i*}\geq c_\textnormal{min} T/n_L, i=1,\dots,n_L  \}$. We have shown $\mathbb{P}(\mathcal{E})\rightarrow 1$. 
	The inequalities below are proved within the set $\mathcal{E}$, and thus hold with probability approaching 1. 
	
	For $i=1,\dots,n_L$, $k=1,\dots,n_L$, $k\neq i$,
	\begin{align*}
	\frac{t_{ik}}{t_k} =\frac{t_{ik}}{\sum_{k'=1}^{n_L} t_{k'k}} \leq \frac{t_{ik}}{t_{ik}+t_{kk}} =\frac{t_{ik}/t_{i*}}{t_{ik}/t_{i*} +t_{kk}/t_{k*} \cdot t_{k*}/t_{i*}} \leq \frac{1}{1+\delta_1\cdot c_{\textnormal{min}}/c_{\textnormal{max}}} =\delta_2 < 1.
	\end{align*}
	Under $H_2$ and $H_3$,  $\min_{i,i'=1,\dots,n_L,i\neq i',j\in V_i} A_{ij} - A_{i'j} = \tau d$, where $\tau$ is bounded away from 0. 
	Now we give a lower bound for $A_{ij}-\bar{A}_{kj}$ for $j \in V_i$ and $k \neq i$,
	\begin{align}
	A_{ij}-\bar{A}_{kj} & = \frac{\sum_t (A_{ij}-P_j^{(t)}) 1(\hat{z}^{(t)}=k)}{t_k} \nonumber \\
	& = \frac{\sum_{k'=1}^{n_L} (A_{ij}-A_{k'j}) t_{k'k}}{t_k} \nonumber \\
	& \geq \frac{\tau d \sum_{k' \neq i} t_{k'k}}{t_k}  \geq	\tau (1-\delta_2) d. \label{eq:Adis}
	\end{align}

	Next, we show the following fact: if $p = \rho_1 d,q = \rho_2 d$ where $\rho_1>\rho_2$ are fixed positive  numbers, then there exists $\delta_3>0$ such that $\textnormal{KL}(p,q)\geq \delta_3 d$, where $\textnormal{KL}(p,q)=p \log \frac{ p}{q}+(1-p) \log \frac{ 1-p}{1-q}$.
	\begin{align*}%\label{eq:kl}
     \textnormal{KL}(p,q) 
     & = p \log \frac{ p}{q} + \log (1- \frac{p-q}{1-q}) + p\log\frac{1-p}{1-q} \\
     & = -p \log \frac{q}{p} +\frac{p-q}{1-q}+o(d) + \rho_1 d \, o(1) \\
     & \geq -p \log \frac{q}{p} + (q-p) + o(d) \\ 
     & = p \left [\frac{q-p}{p} - \log \left (1 + \frac{q-p}{p} \right) \right] + o(d) \\
     & \geq \delta_3 d.
     \end{align*}
 The last line holds for sufficiently small $\delta_3$  because  $\frac{q-p}{p} - \log(1 + \frac{q-p}{p}) = c_{\rho_1,\rho_2}> 0$ where $\frac{q-p}{p} \in (-1,0)$ and $c_{\rho_1,\rho_2}$ is a constant depending on $\rho_1$ and $\rho_2$.
	
	As $\bar{A}_{kj} = \frac{\sum_t P^{(t)}_j 1(z^{(t)}=j)}{t_i} =[\sum_t A_{z^{(t)}_*,j} 1(z^{(t)}=j)]/t_i \asymp d$, combining the above fact and \eqref{eq:Adis}, we have
	\begin{align*}
	L_P(z_*)-L_P(\hat{z}) & = \sum_{t}\sum_{j}\text{KL}(P_j^{(t)},\bar{A}_{\hat{z}^{(t)},j}) \\
	& \geq \sum_{i=1}^{n_L} \sum_{k\neq i} \sum_{t: z_*^{(t)}=i, \hat{z}^{(t)}=k}  \sum_{j \in V_i} \text{KL}(A_{ij},\bar{A}_{\hat{z}^{(t)},j}) \\
	& \geq \sum_{i=1}^{n_L} \sum_{k\neq i} \sum_{t: z_*^{(t)}=i, \hat{z}^{(t)}=k} \sum_{j \in V_i} \tau (1-\delta_2)\delta_3 d\\
	& \geq \tau(1-\delta_2)\delta_3 dvnT_e/n_L.
	\end{align*}
	Letting $\delta = 1/[\tau(1-\delta_2)\delta_3]$, 
$$
\frac{\delta n_L}{dvnT}(L_P(z_*)-L_P(\hat{z})) \geq \frac{T_e}{T},
$$	
with probability approaching 1.
\end{proof}

To prove Theorem 3, we need the following lemma.

\begin{manuallemma}{S1}\label{lem:lem1}
\begin{align*}
&\mathbb{P}(\max_z |L_G(z) - L_P(z)| \geq 2\eta) \leq \\
& \quad \ n_L^T  (T/n_L+1)^{n_L n} e^{-\eta} 
+ 2n_L^T \exp \left \{-\frac{\eta^2/4}{\sum_t\sum_j (\log \bar{A}_{ij})^2\textnormal{Var}(G_j^{(t)})+\max_{ij} |\log\bar{A}_{ij}|\eta/6} \right\} \\
& \quad + 2n_L^T\exp \left \{-\frac{\eta^2/4}{\sum_{ij:\bar{A}_{ij}<1} \left ( (\log(1-\bar{A}_{ij}))^2  \sum_{t:z^{(t)}=i} \textnormal{Var}(G_j^{(t)}) \right ) 
 + \max_{ij:\bar{A}_{ij}<1}|\log (1-\bar{A}_{ij})|\eta/6} \right \}.
\end{align*}
\end{manuallemma}

\begin{proof}[Proof of Lemma S1]
\begin{align*}
	%\begin{split}
	L_G(z)-L_P(z)  = &  \left ( \sum_{i=1}^{n_L} t_i \sum_{j } \hat{A}_{ij} \log \hat{A}_{ij}+(1-\hat{A}_{ij}) \log (1-\hat{A}_{ij}) \right ) \nonumber \\
	& -  \left ( \sum_{i=1}^{n_L} t_i \sum_{j } \hat{A}_{ij} \log \bar{A}_{ij}+(1-\hat{A}_{ij}) \log (1-\bar{A}_{ij})  \right ) \nonumber \\
	& + \left ( \sum_{i=1}^{n_L} t_i \sum_{j } \hat{A}_{ij} \log \bar{A}_{ij}+(1-\hat{A}_{ij}) \log (1-\bar{A}_{ij})  \right ) \nonumber \\
	& - \left ( \sum_{i=1}^{n_L} t_i \sum_{j } \bar{A}_{ij} \log \bar{A}_{ij}+(1-\bar{A}_{ij}) \log (1-\bar{A}_{ij})  \right ) \nonumber \\
	= & \sum_{i=1}^{n_L} t_i \sum_{j } D(\hat{A}_{ij}|\bar{A}_{ij}) +B_{n_L,n,T}. %\label{eq:Lpg}
	% \end{split}
\end{align*}
%We  bound $D(\hat{A}_{ij}|\bar{A}_{ij})$ and $B_{n_L,n,T}$ separately. 
To bound $\sum_{i=1}^{n_L} t_i \sum_{j} D(\hat{A}_{ij}|\bar{A}_{ij})$, 
we adopt the approach in \cite{Choietal2011}, which is based on a heterogeneous Chernoff bound in \cite{dubhashi2009concentration}. 
Let $\nu$ be any realization of $\hat{A}$. 	
\begin{align*}
	\mathbb{P}(\hat{A}_{ij}=\nu_{ij}|z_{*})\leq e^{-t_i D(\nu_{ij} | \bar{A}_{ij})}.
	\end{align*}
	By the independence of $\hat{A}_{ij}$ conditional on $z_{*}$,
	\begin{align*}
	\mathbb{P}(\hat{A}=\nu|z_*) \leq \exp \left \{ -\sum_{i=1}^{n_L} \sum_{j} t_i D( \nu_{ij} |\bar{A}_{ij}) \right \}.
\end{align*}
Let $\hat{\mathcal{A}}$ be the range of $\hat{A}$ for a fixed $z$. Then $|\hat{\mathcal{A}} |\leq \prod_{i=1}^{n_L} (t_i+1)^n \leq  \prod_{i=1}^{n_L} (t_i+1)^n  \leq (T/n_L+1)^{n_L n}$, as $\hat{A}_{ij}$ can only take values from $0/t_i,1/t_i,\dots,t_i/t_i$.
    
For all $\eta>0$,
\begin{align*}
& \mathbb{P} \left ( \left . \sum_{i=1}^{n_L} \sum_{j } t_i D(\hat{A}_{ij} |\bar{A}_{ij}) \geq \eta  \right | z_* \right ) \\
=& \sum_{\nu \in \hat{\mathcal{A}}} \mathbb{P} \left ( \left . \hat{A}=\nu, \sum_{i=1}^{n_L} \sum_{j } t_i D(\nu_{ij} |\bar{A}_{ij}) \geq \eta \right | z_* \right ) \\
\leq & \sum_{\nu \in \hat{\mathcal{A}}} \exp \left \{ -\sum_{i=1}^{n_L} \sum_{j } t_i D( 
\nu_{ij} |\bar{A}_{ij}) \right \} 1 \left \{  -\sum_{i=1}^{n_L} \sum_{j } t_i D(\nu_{ij} |\bar{A}_{ij}) \leq - \eta \right \} \\
\leq  & \sum_{\nu \in \hat{\mathcal{A}}} e^{-\eta} \leq  | \hat{\mathcal{A}} |  e^{-\eta} \leq  (T/n_L+1)^{n_L n} e^{-\eta}, 
\end{align*}
and then
\begin{align}
\mathbb{P} \left (  \max_{z} \sum_{i=1}^{n_L} \sum_{j } t_i D( \hat{A} |\bar{A}_{ij}) \geq \eta \right ) \leq n_L^T  (T/n_L+1)^{n_L n} e^{-\eta}. \label{eq:KL}
\end{align}

Next, we bound $B_{n_L,n,T}$. 
%\begin{align*}
%	B_{n_L,n,T} &= \sum_{i} t_i \left(\sum_{j}(\hat{A}_{ij}\log\bar{A}_{ij} + 
%	(1-\hat{A}_{ij})\log(1-\bar{A}_{ij})) - \sum_{j}(\bar{A}_{ij}\log\bar{A}_{ij} +
%	(1-\bar{A}_{ij})\log(1-\bar{A}_{ij}))  \right)	\\
%	&= \sum_i t_i (\sum_j (\hat{A}_{ij}-\bar{A}_{ij})\log\bar{A}_{ij}) + 
%	\sum_i t_i (\sum_j (\bar{A}_{ij}-\hat{A}_{ij})\log(1-\bar{A}_{ij}))\\
%	&= \sum_i\left(\sum_j \sum_{t:z^{(t)}=i} (G_j^{(t)}-P_j^{(t)})\log\bar{A}_{ij} \right) + 
%	 \sum_i\left(\sum_j \sum_{t:z^{(t)}=i} (G_j^{(t)}-P_j^{(t)})\log(1-\bar{A}_{ij}) \right).
%\end{align*}
Let $B_{n_L,n,T} = B_{1,n_L,n,T} + B_{2,n_L,n,T}$, where
\begin{align*}
    B_{1,n_L,n,T} &= \sum_i\left(\sum_j \sum_{t:z^{(t)}=i} (G_j^{(t)}-P_j^{(t)})\log\bar{A}_{ij} \right),  \\
    B_{2,n_L,n,T} &= \sum_i\left(\sum_j \sum_{t:z^{(t)}=i} (G_j^{(t)}-P_j^{(t)})\log(1-\bar{A}_{ij}) \right).
\end{align*}
As $\left |(G_j^{(t)}-P_j^{(t)})\log\bar{A}_{ij} \right| \leq |\log\bar{A}_{ij}|$, 
by Bernstein's inequality, we have 
\begin{align} 
    &\mathbb{P}(|B_{1,n_L,n,T}| \geq \eta/2) \leq 2\exp \left \{-\frac{\eta^2/4}{\sum_t\sum_j (\log \bar{A}_{ij})^2\textnormal{Var}(G_j^{(t)})+\max_{ij} |\log\bar{A}_{ij}|\eta/6} \right\}, \nonumber \\
    &\mathbb{P}(\max_z |B_{1,n_L,n,T}| \geq \eta/2) \leq 2n_L^T\exp \left \{-\frac{\eta^2/4}{\sum_t\sum_j (\log \bar{A}_{ij})^2\textnormal{Var}(G_j^{(t)})+\max_{ij} |\log\bar{A}_{ij}|\eta/6} \right\}. \label{eq:B1_ub}
\end{align}
In addition, if $\bar{A}_{ij}=1$, $\sum_{t:z^{(t)}=i} (G_j^{(t)}-P_j^{(t)}) \equiv 0$, which implies the term $\sum_{t:z^{(t)}=i} (G_j^{(t)}-P_j^{(t)}) \log(1-\bar{A}_{ij})$ in $B_{2,n_L,n,T}$ can be dropped.
As $\left |(G_j^{(t)}-P_j^{(t)})\log (1-\bar{A}_{ij}) \right| \leq |\log (1-\bar{A}_{ij})|$, by Bernstein's inequality,
\begin{align}
	&\mathbb{P} (|B_{2,n_L,n,T}| \geq \eta/2)  \nonumber \\
	& \leq 2\exp \left \{-\frac{\eta^2/4}{\sum_{ij:\bar{A}_{ij}<1} \left ( (\log(1-\bar{A}_{ij}))^2  \sum_{t:z^{(t)}=i} \textnormal{Var}(G_j^{(t)}) \right )  + \max_{ij:\bar{A}_{ij}<1}|\log (1-\bar{A}_{ij})|\eta/6} \right \}, \nonumber \\
	&\mathbb{P} (\max_z |B_{2,n_L,n,T}| \geq \eta/2)  \nonumber \\
	& \leq 2n_L^T \exp \left \{-\frac{\eta^2/4}{\sum_{ij:\bar{A}_{ij}<1} \left ( (\log(1-\bar{A}_{ij}))^2  \sum_{t:z^{(t)}=i} \textnormal{Var}(G_j^{(t)}) \right )  + \max_{ij:\bar{A}_{ij}<1}|\log (1-\bar{A}_{ij})|\eta/6} \right \}.\label{eq:B2_ub}
\end{align}
	
Finally, combining \eqref{eq:KL}, \eqref{eq:B1_ub} and \eqref{eq:B2_ub}, we obtain
\begin{align*}
&\mathbb{P}(\max_z |L_G(z) - L_P(z)| \geq 2\eta)  \\
& \quad \leq \mathbb{P} \left (  \max_{z} \sum_{i=1}^{n_L} \sum_{j } t_i D( \hat{A} |\bar{A}_{ij}) \geq \eta \right ) + \mathbb{P}(\max_z |B_{1,n_L,n,T}|\geq \eta/2) + \mathbb{P}(\max_z |B_{2,n_L,n,T}|\geq \eta/2)\\
& \quad \leq n_L^T  (T/n_L+1)^{n_L n} e^{-\eta} 
+ 2n_L^T \exp \left \{-\frac{\eta^2/4}{\sum_t\sum_j (\log \bar{A}_{ij})^2\textnormal{Var}(G_j^{(t)})+\max_{ij} |\log\bar{A}_{ij}|\eta/6} \right\} \\
& \quad\  + 2n_L^T \exp \left \{-\frac{\eta^2/4}{\sum_{ij:\bar{A}_{ij}<1} \left ( (\log(1-\bar{A}_{ij}))^2  \sum_{t:z^{(t)}=i} \textnormal{Var}(G_j^{(t)}) \right )
 + \max_{ij:\bar{A}_{ij}<1}|\log (1-\bar{A}_{ij})|\eta/6} \right \}.
\end{align*}
\end{proof}

%%%%%%%%%%%%%%%%%%%%%%%%%-------------------------Theorem 3
\begin{proof}[Proof of Theorem 3]

First we show the following fact: under $H_1 - H_4$,  if $n_L^2\log T/(dTv) \rightarrow 0$, $(\log d)^2n_L^2\log n_L/(dnv^2) \rightarrow 0$ and $(\log T)^2n_L^2\log n_L/(dnv^2) \rightarrow 0$, then
\begin{align}
\max_z \frac{n_L}{dvnT}|L_P(z) - L_G(z)| = o_p(1), \quad \textnormal{as}\ n_L\rightarrow \infty, n \rightarrow \infty, T\rightarrow \infty. \label{fact1}
\end{align}
Letting $\eta = dvnT \epsilon/n_L$, the LHS in Lemma \ref{lem:lem1} becomes
$\mathbb{P}(\max_z \frac{n_L}{dvnT}|L_G(z) - L_P(z)| \geq 2\epsilon)$.
To prove the above fact, we need to show each term in the RHS of Lemma \ref{lem:lem1} goes to 0.

For the first term, it is easy to check that if $n_L\log n_L/(dvn)\rightarrow 0$ and $n_L^2\log T/(dvT) \rightarrow 0$, then 
\begin{align*}
    n_L^T(T/n_L)^{n_L n}e^{-\frac{dvnT\epsilon}{n_L}} \rightarrow 0.
\end{align*}
Under $H_2$, $A_{ij} \asymp d$ and  $|\log\bar{A}_{ij}| =O (|\log d|)$ for $i\neq j$. We can therefore find a constant $C_1$ such that
\begin{align*}
\mathbb{P}(|B_{1,n_L,n,T}| \geq dvnT\epsilon/(2n_L)) \leq 2\exp \left \{-\frac{d^2v^2n^2T^2\epsilon^2/(4n_L^2)}{C_1^2 Tn(\log d)^2d + C_1 |\log d| dvnT\epsilon/(6n_L)} \right\},
\end{align*}
and
\begin{align*}
\mathbb{P}(\max_z |B_{1,n_L,n,T}| \geq dvnT\epsilon/(2n_L)) \leq  2n_L^T \exp \left \{-\frac{d^2v^2n^2T^2\epsilon^2/(4n_L^2)}{C_1^2Tn(\log d)^2d + C_1|\log d|dvnT\epsilon/(6n_L)} \right \}. 
\end{align*}
Then if $(\log d)^2 n_L^2\log n_L/(dnv^2) \rightarrow 0$,
\begin{align*}
\mathbb{P}(\max_z |B_{1,n_L,n,T}| \geq dvnT\epsilon/(2n_L)) \rightarrow 0.
\end{align*}
For the third term, when $\bar{A}_{ij} <1$,  we have
%\tcb{The following inequalities are the reason for the weird condition $(\log T)^2/n \rightarrow 0$.
%My understanding is since we are considering $\bar{A}_{ij} < 1$, then $1 - \bar{A}_{ij}$ is a constant and $|\log(1-\bar{A}_{ij})|$ is also a constant.}
\begin{align*}
\bar{A}_{ij} &\leq \frac{(t_i-1) + P_j^{(t)}}{t_i}, \\
1-\bar{A}_{ij} &\geq \frac{1-P_j^{(t)}}{t_i} \geq \frac{1-P_j^{(t)}}{T},
\end{align*}
which imply $|\log(1-\bar{A}_{ij})| \leq C_2\log T$ for some constant $C_2>0$.
Therefore,
\begin{align*}
\mathbb{P} (\max_z |B_{2,n_L,n,T}| \geq dvnT\epsilon/(2n_L)) \leq 2n_L^T\exp \left \{-\frac{d^2v^2n^2T^2 \epsilon^2  /(4n_L^2)}{C_2^2(\log T)^2 Tnd +C_2 (\log T) dnvT \epsilon /(6n_L)} \right \}. 
\end{align*}
Furthermore, if $(\log T)^2 n_L^2\log n_L/(dnv^2) \rightarrow 0$,
\begin{align*}
\mathbb{P}(\max_z |B_{2,n_L,n,T}| \geq dvnT\epsilon/(2n_L)) \rightarrow 0.
\end{align*}
Combining the inequalities of the above three terms, we have proved \eqref{fact1}.

Finally, for all $\epsilon>0$, 
	\begin{align*}
	\mathbb{P} \left ( \frac{T_e}{T} \geq \epsilon  \right ) 
	= & \mathbb{P} \left ( \frac{T_e}{T} \geq \epsilon, \frac{\delta  n_L }{d v nT}(L_P(z_*)-L_P(\hat{z})) \geq \frac{T_e}{T} \right ) \\
	&  +  \mathbb{P} \left ( \frac{T_e}{T} \geq \epsilon,  \frac{\delta  n_L }{d v nT}(L_P(z_*)-L_P(\hat{z})) < \frac{T_e}{T}  \right )\\
	= & \mathbb{P} \left (  \frac{\delta  n_L }{d v nT}(L_P(z_*)-L_P(\hat{z}))  \geq \epsilon \right )+ o(1) \quad \textnormal{(by Lemma 1)} \\
	= & \mathbb{P} \left (  \frac{\delta  n_L }{d v nT} \left[ (L_P(z_*)-L_G(z_*))+ (L_G(z_*)-L_G(\hat{z}))+(L_G(\hat{z})-L_P(\hat{z})) \right] \geq \epsilon \right )+ o(1) \\
	\leq & \mathbb{P} \left (  \frac{\delta  n_L }{d v nT} \left ( |L_P(z_*)-L_G(z_*)|+|L_G(\hat{z})-L_P(\hat{z})| \right ) \geq \epsilon \right )+ o(1) \\
	\rightarrow &\  0.
	\end{align*}
\end{proof}

We now give the result of label consistency for fixed $n_L$. 
We make the following assumptions similar to $H_1$ -- $H_4$.
\begin{itemize}
    \item [$H_1'$:] $c_{\textnormal{min}} T \leq t_{i*} \leq c_{\textnormal{max}} T$ for $i=1,\dots,n_L$.

    \item [$H_2'$:] $A_{ij} = s_{ij}d$ for $i=1,\dots,n_L$,$j=1,\dots,n$ and $i\neq j$ where $s_{ij}$ are unknown constants satisfying $0<s_{\textnormal{min}} \leq s_{ij} \leq s_{\textnormal{max}}<\infty$  while $d$ goes to 0 as $n$ goes to infinity.
    \item [$H_3'$:] There exists a set $V_i \subset \{n_L + 1, \dots,n\}$ for $i = 1,\dots,n_L$ with $|V_i| \geq vn$ such that $\tau = \min_{i,i' = 1,\dots,n_L,i\neq i',j\in V_i} (s_{ij}-s_{i'j})$ is bounded away from 0.
    \item [$H_4'$:] $A_{ii'}$ is bounded away from 1 for $i=1,\dots,n_L$,$i'=1,\dots,n_L$ and $i\neq i'$.
\end{itemize}
%%%%%
\begin{manualtheorem}{3$'$}\label{thm:thm3'}
Under $H_1' - H_4'$, if  $\log T/(dTv) = o(1)$, $(\log d)^2/(dnv^2) = o(1)$ and 
$(\log T)^2/(dnv^2) = o(1)$, then
\begin{align*}
    T_e/T = o_p(1), \quad \textnormal{as} \  n \rightarrow \infty, T \rightarrow \infty.
\end{align*}
\end{manualtheorem}

We omit all the proofs for fixed $n_L$ because they are trivial corollaries of the results for growing $n_L$.

%%%%%%%%%%%%%%%%%%%%%%%---------------------Theorem 4
\begin{proof}[Proof of Theorem 4]

   First we show the following fact: under the conditions in Theorem 4,
   	\begin{align*}
	n_L T_e/T = o_p(1), \,\, \textnormal{as} \ n_L \rightarrow \infty, n\rightarrow \infty, T \rightarrow \infty.
	\end{align*}
	According to the proof in Theorem 3, we need 
	$$
	\mathbb{P} \left (  \frac{\delta  n_L^2 }{d v nT} \left ( |L_P(z_*)-L_G(z_*)|+|L_G(\hat{z})-L_P(\hat{z})| \right ) \geq \epsilon \right ) \rightarrow 0,
	$$
	which holds if we can show
	$$
	\max_z \frac{  n_L^2 }{d v nT}|L_G(z)-L_P(z)| =o_p(1).
	$$
	As in the proof of Lemma \ref{lem:lem1}, this holds by letting $\eta = d v n T\epsilon /n_L^2$.

	Then we bound $\left |\hat{A}_{ij}^{\hat{z}}-\hat{A}_{ij}^{z_*} \right|$:
	\begin{align*}
	&|\hat{A}_{ij}^{\hat{z}}-\hat{A}_{ij}^{z_*}| 
	=  \left |\frac{\sum_t G_j^{(t)} 1 (\hat{z}^{(t)}=i) }{  t_i}-\frac{\sum_t G_j^{(t)} 1 (z_*^{(t)}=i) }{  t_{i*}}  \right | \\
    &\leq  \left |\frac{\sum_t G_j^{(t)} 1 (\hat{z}^{(t)}=i) }{  t_i}-\frac{\sum_t G_j^{(t)} 1 (\hat{z}^{(t)}=i) }{  t_{i*}}  \right | 
	 +\left |\frac{\sum_t G_j^{(t)} 1 (\hat{z}^{(t)}=i) }{  t_{i*}}-\frac{\sum_t G_j^{(t)} 1 (z_*^{(t)}=i) }{  t_{i*}}  \right | \\
	&\leq  \left | \frac{t_{i*}-t_{i}}{t_{i*}}  \right |+\frac{\sum_{t} \left | 1 (\hat{z}^{(t)}=i)- 1 (z_*^{(t)}=i) \right |}{t_{i*}} 
	\leq  \delta n_L T_e/T,
	\end{align*}
	where $\delta$ is a constant. The last line holds by $H_1'$.
	
	Furthermore,
	\begin{align*}
	& \mathbb{P} \left (\max_{ij} \left|\hat{A}_{ij}^{\hat{z}}-A_{ij} \right| \geq \epsilon \right ) \\
	\leq & \mathbb{P} \left (\max_{ij} \left|\hat{A}_{ij}^{\hat{z}}-\hat{A}_{ij}^{z_*}  \right | \geq \epsilon/2 \right)+P \left (\max_{ij} \left|\hat{A}_{ij}^{z_*}-A_{ij}  \right | \geq \epsilon/2 \right) \\
	\leq & \mathbb{P} \left (\delta n_L T_e/T \geq \epsilon \right )+P \left (\max_{ij} \left|\hat{A}_{ij}^{z_*}-A_{ij}  \right | \geq \epsilon/2 \right).
	\end{align*}
The second term vanishes by Hoeffding's inequality: for all $\epsilon>0$,
	\begin{align*}
	&\mathbb{P} \left ( \left . \left|\hat{A}_{ij}^{z_*}-A_{ij}  \right | \geq \epsilon/2  \right |z_* \right) \\
	= &  \mathbb{P} \left (  \left . \left | \sum_t 1 (z_{*}^{(t)}=i) (G_j^{(t)}-A_{ij}) \right | \geq \epsilon t_{i*}/2 \right | z_* \right) \\
	\leq & 2\exp \{-\epsilon^2 t_{i*}/2 \}.
	\end{align*}
	Therefore, 	if $n_L\log n/T \rightarrow 0$,
	\begin{align*}
	& \mathbb{P} \left (\max_{ij} \left|\hat{A}_{ij}^{z_*}-A_{ij} \right| \geq \epsilon/2 \right ) \\
	\leq & 2 n n_L  \exp \{- \epsilon^2 c_{\textnormal{min}} T/(2n_L) \} + \mathbb{P}(t_{i*} < c_\textnormal{min} T/n_L, \textnormal{for some $i$}) \rightarrow 0.
	\end{align*}
\end{proof}

The following theorem is on estimation consistency for fixed $n$.
\begin{manualtheorem}{4$'$}\label{thm:thm4'}
Under $H_1' - H_4'$, if $\log n/T = o(1)$, $\log T/(dTv) = o(1)$, $(\log d)^2/(dnv^2) = o(1)$ and 
$(\log T)^2/(dnv^2) = o(1)$, then
\begin{align*}
    \max_{i\in \{1,\dots,n_L\},j\in\{1,\dots,n\}} \left|\hat{A}_{ij}^{z_*}-A_{ij} \right| = o_p(1), \quad \textnormal{as }  n \rightarrow \infty, T \rightarrow \infty.
\end{align*}
\end{manualtheorem}

Finally, we give the simplest version of the estimation consistency result, which only considers the rates of $n$ and $T$ but treats $n_L$, $d$, and $v$ as fixed. 
\begin{manualtheorem}{4$''$}\label{thm:thm4''}
Under $H_1' - H_4'$, for fixed $d$ and $v$, if $\log n/T = o(1)$ and 
$(\log T)^2/n = o(1)$, then
\begin{align*}
    \max_{i\in \{1,\dots,n_L\},j\in\{1,\dots,n\}} \left|\hat{A}_{ij}^{z_*}-A_{ij} \right| = o_p(1), \quad \textnormal{as } n \rightarrow \infty, T \rightarrow \infty.
\end{align*}
\end{manualtheorem}
The first condition means $n$ can grow faster than $T$ as long as $\log n/T \rightarrow 0$. Such a condition is common in the literature of high-dimensional statistics. The second condition is more of a technical one: for proving the label consistency, we need an upper bound of the growth rate of $T$ due to the concentration bound in Lemma \ref{lem:lem1}. 

%%%%%%%%%%%%%%%%%%%%%%%%%%%%%%%-------------------Lemma 2
\begin{proof}[Proof of Lemma 2] 
	By the proof of Lemma 1,  there exists $\delta_1>0$ such that  
	\begin{align}
	& t_{ii}+t_{i0} \geq \delta_1 t_{i*},   \quad i =1,\dots,n_L, \label{tricky}\\
	& t_{00} \geq \delta_1 t_{0*},  \label{t00} 
	\end{align}
	with probability approaching 1.
	
	Therefore\footnote{Some inequalities below hold with probability approaching 1. We omit this sentence occasionally.}, for $i=1,\dots,n_L$, $j\in V_i$,
	\begin{align*}
	A_{ij}-\bar{A}_{0j} & = \frac{\sum_t (A_{ij}-P_j^{(t)}) 1(\hat{z}^{(t)}=0)}{t_0} \\
	& = \frac{\sum_{k=0}^{n_L} (A_{ij}-A_{kj}) t_{k0}}{t_0} \\
	& \geq \frac{(A_{ij}-A_{0j})t_{00}}{t_0}	\geq \tau d \frac{t_{00}}{T} \geq \tau d \frac{t_{00}}{(n_L+1)t_{0*}/c_{\textnormal{min}}} \geq  \frac{\tau dc_{\textnormal{min}} \delta_1}{n_L}.
	\end{align*}
	
	Using the same argument in Lemma 1, it follows that
	\begin{align}
	L_P(z_*)-L_P(\hat{z})  = &\sum_{t}\sum_{j} \textnormal{KL}(P_j^{(t)},\bar{A}_{\hat{z}^{(t)},j}) \nonumber \\
	\geq & \max _{i = 1,\dots,n_L} \sum_{t:z_*^{(t)}=i,\hat{z}^{(t)}=0}\sum_{j \in V_i} \textnormal{KL}(A_{ij},\bar{A}_{0j}) \nonumber\\
	\geq & \max_{i = 1,\dots,n_L} \frac{\tau dc_{\min} \delta_1 \delta_3}{n_L}\frac{vn}{n_L}t_{i0} \nonumber \\
	\geq & \max_{i=1,\dots,n_L} \frac{\tau dc_{\min} \delta_1 \delta_3}{n_L} \frac{vn}{n_L} \frac{t_{i0}}{t_{i*}} \frac{c_{\min} T}{n_L} \nonumber \\
	\geq & \max_{i=1,\dots,n_L} \tau \epsilon \frac{dvnT}{n_L^3}\frac{t_{i0}}{t_{i*}}, \label{thm:separate_temp}
	\end{align}
	where $\epsilon$ is a positive constant and $\tau$ is bounded away from 0. 
	
	Next, we show the following fact: under the conditions in Lemma 2,
    \begin{align*}
	\max_z \frac{  n_L^3 }{dvnT}|L_G(z)-L_P(z)| =o_p(1).
	\end{align*}
	As in the proofs of Lemma \ref{lem:lem1} and Theorem 3, the above statement holds by letting $\eta = d v n T\epsilon / n_L^3$. Combining \eqref{thm:separate_temp} and the above fact, by the same argument in Theorem 3, we have
	\begin{align}
	\mathbb{P}\left ( \max_{i=1,\dots,n_L}\frac{t_{i0}}{t_{i*}} \leq \eta \right ) \rightarrow 1. \label{tricky_solved}
	\end{align}
\end{proof}

%%%%%%%%%%%%%%%%%%%%%%%%%%%---------------------Theorem 5
\begin{proof}[Proof of Theorem 5]
	Due to \eqref{tricky} and \eqref{tricky_solved}, there exists $\delta_2>0$ such that 
	\begin{align*}
	t_{ii}\geq \delta_2 t_{i*}   \quad \mbox{for } i =0,\dots,n_L,
	\end{align*}
	with probability approaching 1. By the same argument in Lemma 1,
	\begin{align*}
	L_P(z_*)-L_P(\hat{z})  = & \sum_{t=1}^T \sum_{j=1}^n \textnormal{KL}(P_j^{(t)}, \bar{A}_{\hat{z}^{(t)},j}) \\
	\geq & \sum_{i=1}^{n_L} \sum_{0\leq k \leq n_L, k\neq i} \sum_{t: z_*^{(t)}=i, \hat{z}^{(t)}=k}  \sum_{j \in V_i} \textnormal{KL}(A_{ij},\bar{A}_{kj}) \\
	\geq &   \frac{vn}{n_L}  \sum_{i=1}^{n_L} \sum_{0\leq k \leq n_L, k\neq i} t_{ik} \tau (1-\delta_2)\delta_3 d,
	\end{align*}
	which implies that there exists $\delta>0$ such that with probability approaching 1,
	\begin{align}\label{sep11}
	\frac{\delta n_L}{d v n T } (L_P(z_*)-L_P(\hat{z})) \geq  \sum_{i=1}^{n_L} \sum_{0\leq k \leq n_L, k\neq i} \frac{ t_{ik}}{T}.
	\end{align}
	By the same argument in Theorem 3, this further implies 
	\begin{align}
	\sum_{i=1}^{n_L} \sum_{0\leq k \leq n_L, k\neq i} \frac{ t_{ik}}{T} =o_p(1), \quad \mbox{as } n_L\rightarrow \infty, n\rightarrow \infty, T\rightarrow \infty, \label{part1}
	\end{align}
	if $ n_L^2\log T/(d v T)=o(1) $, $n_L^2 (\log T)^2 \log n_L/ (dnv^2)=o(1)$ and $n_L^2 (\log d)^2 \log n_L/ (dnv^2)=o(1)$. 
	
	As in the proof of Theorem 4,
	\begin{align}
	\sum_{i=1}^{n_L} \sum_{0\leq k \leq n_L, k\neq i} (n_L+1)\frac{t_{ik}}{T} =o_p(1), \quad \mbox{as } n_L\rightarrow \infty, n\rightarrow \infty, T\rightarrow \infty, \label{novel} 
	\end{align}
	if $n_L^3\log T/(d v T)=o(1)$, $n_L^4 (\log T)^2\log n_L / (dnv^2)=o(1)$ and $n_L^4 (\log d)^2\log n_L / (dnv^2)=o(1)$. 
	
	Now we bound $t_{0i}$, $i=1,\dots,n_L$.     From \eqref{novel},
	$\sum_{1\leq k\leq n_L,k\neq i} t_{ki} = o_p( T/(n_L+1))$. And from $\delta_2 Tc_{\textnormal{min}}/(n_L+1) \leq  \delta_2 t_{i*} \leq t_{ii}$, $\sum_{1\leq k\leq n_L,k\neq i} t_{ki} \leq t_{ii}$, with probability approaching 1. Moreover, from \eqref{t00}, $t_{0i}\leq (1-\delta_1) t_{0*}$.
	
	Therefore,  there exists $\delta_4>0$ such that for $i=1,\dots,n_L$, $j\in V_i$,
	\begin{align*}
	A_{ij}-\bar{A}_{ij} & = \frac{\sum_t (A_{ij}-P_j^{(t)}) 1(\hat{z}^{(t)}=i)}{t_i} \\
	& \geq \frac{(A_{ij}-A_{0j})t_{0i}}{t_i} \\
	& \geq \frac{\tau dt_{0i}}{t_{0i}+t_{ii}+\sum_{1\leq k\leq n_L,k\neq i} t_{ki}} \\
	& \geq \frac{\tau dt_{0i}}{(1-\delta_1)t_{0*}+2 t_{ii}} \\
	& \geq \frac{\tau dt_{0i}}{(1-\delta_1)t_{0*}+2 t_{i*}} \geq \frac{\tau d n_L t_{0i}}{\delta_4 T}.
	\end{align*}
	It follows that
	\begin{align}
	L_P(z_*)-L_P(\hat{z})  \geq & \max_{i=1,\dots,n_L} \sum_{t: z_*^{(t)}=i, \hat{z}^{(t)}=i}  \sum_{j \in V_i} \textnormal{KL}(A_{ij},\bar{A}_{ij})  \nonumber \\
	\geq & \max_{i=1,\dots,n_L}   \frac{\tau d n_L t_{0i}\delta_3}{\delta_4 T} \frac{vn}{n_L+1} t_{ii} \nonumber  \\
	\geq & \max_{i=1,\dots,n_L}   \frac{d}{\delta_4}  \frac{n_L t_{0i}}{T} \frac{vn}{n_L+1}  \tau \delta_2\delta_3 t_{i*} \nonumber \\
	\geq & \max_{i=1,\dots,n_L} \frac{d}{\delta_4} \frac{n_L t_{0i}}{T} \frac{vn}{n_L+1}  \tau\delta_2\delta_3 T \frac{c_{\min}}{n_L+1} \nonumber  \\
	\geq & \max_{i=1,\dots,n_L} \frac{ d v nT}{n_L^2} \frac{n_L t_{0i}}{T}\delta,  \label{sep22}
	\end{align}
	where $\delta = \tau \delta_2\delta_3 c_{min}/\delta_4$ is positive constant. 
	
	By using the same argument in Theorem 3, 
	\begin{align}
	%\mathbb{P}\left ( \max_{i=1,\dots,n_L}\frac{t_{0i}}{t_{i*}} \geq \eta \right ) \rightarrow 0
	\max_{i=1,\dots,n_L}  \frac{n_L t_{0i}}{T} =o_p(1),  \label{part2}
	\end{align}
	if $ n_L^4\log T/(d v T)=o(1) $, $ (\log T)^2 n_L^6\log n_L/ (d n v^2)=o(1) $ and $n_L^6 (\log d)^2\log n_L / (dnv^2)=o(1)$.  It follows that
	\begin{align*}
	\sum_{i=1}^{n_L} \frac{t_{0i}}{T}=o_p(1).
	\end{align*}
	Combining \eqref{part1} and \eqref{part2},
	\begin{align*}
	\frac{T_e}{T} =o_p(1), \quad \mbox{as }  n\rightarrow \infty, T\rightarrow \infty.
	\end{align*}	
\end{proof}

For label consistency under the hub model with the null component with fixed $n_L$, we make the following assumptions:
\begin{itemize}
    \item [$H_1^{*\prime}$:] $Tc_{\textnormal{min}}/{n_L} \leq t_{i*} \leq Tc_{\textnormal{max}}/{n_L}$ for $i=0,\dots,n_L$.

    \item [$H_2^{*\prime}$:] $A_{ij} = s_{ij}d$ for $i=0,\dots,n_L$,$j=1,\dots,n$ and $i\neq j$ where $s_{ij}$ are unknown constants satisfying $0<s_{\textnormal{min}} \leq s_{ij} \leq s_{\textnormal{max}}<\infty$ while $d$ goes to 0 as $n$ goes to infinity.
    \item [$H_3^{*\prime}$:] There exists a set $V_i \subset \{n_L + 1, \dots,n\}$ for $i = 1,\dots,n_L$ with $|V_i| \geq vn$ such that $\tau = \min_{i = 1,\dots,n_L,i' = 0,\dots,n_L,i\neq i',j\in V_i} (s_{ij}-s_{i'j})$ is bounded away from 0.
    \item [$H_4^{*\prime}$:] $A_{ii'}$ is bounded away from 1 for $i=0,\dots,n_L$,$i'=1,\dots,n_L$ and $i\neq i'$.
\end{itemize}
%%%%%
\begin{manualtheorem}{5$'$}\label{thm:thm5'}
Under $H_1^{*\prime} - H_4^{*\prime}$, if  $\log T/(dTv) = o(1)$, $(\log d)^2/(dnv^2) = o(1)$ and 
$(\log T)^2/(dnv^2) = o(1)$, then
\begin{align*}
    T_e/T = o_p(1), \quad \textnormal{as} \ n \rightarrow \infty, T \rightarrow \infty.
\end{align*}
\end{manualtheorem}
%%%%%%%%%%%%%%%%%%%%%%%%%%%%--------------------Theorem 6
\begin{proof}[Proof of Theorem 6]
	By the same argument in Theorem 4, it is sufficient to show  
	\begin{align}\label{final}
	\frac{(n_L+1) T_e}{T} =o_p(1), \quad \mbox{as } n_L\rightarrow \infty, n\rightarrow \infty, T\rightarrow \infty.
	\end{align}	
	From \eqref{novel}, we have shown 
	\begin{align*}
	\sum_{i=1}^{n_L} \sum_{0\leq k \leq n_L, k\neq i} \frac{ (n_L+1) t_{ik}}{T} =o_p(1), \quad \mbox{as } n_L\rightarrow \infty, n\rightarrow \infty, T\rightarrow \infty.
	\end{align*}
	From \eqref{sep22}, there exists $\delta'>0$ such that
	\begin{align*}
	L_P(z_*)-L_P(\hat{z})  \geq    \max_{i=1,\dots,n_L} \frac{ d v nT}{\delta' n_L^3}  \frac{n_L(n_L+1) t_{0i}}{T},
	\end{align*}
	which further implies
	\begin{align*}
	\max_{i=1,\dots,n_L}  \frac{n_L(n_L+1)t_{0i}}{T} =o_p(1), 
	\end{align*}
	if $ n_L^5\log T/(dTv)=o(1) $, $(\log d)^2 n_L^8\log n_L/(dnv^2) = o(1)$ and $ (\log T)^2 n_L^8\log n_L/ (d n v^2)=o(1) $.
	
	It follows that
	\begin{align*}
	\sum_{i=1}^{n_L} \frac{(n_L+1)t_{0i}}{T}=o_p(1).
	\end{align*}
	Eq. \eqref{final} is therefore proved and so is the theorem. 
\end{proof}

Finally, we give the result for estimation consistency under the hub model with the null component with fixed $n_L$:
\begin{manualtheorem}{6$'$}\label{thm:thm6'}
Under $H_1^{*\prime} - H_4^{*\prime}$, if  $\log n/T = o(1)$, $\log T/(dTv) = o(1)$, $\log T/(dTv) = o(1)$, $(\log d)^2/(dnv^2) = o(1)$ and 
$(\log T)^2/(dnv^2) = o(1)$, then
\begin{align*}
    \max_{i\in \{0,\dots,n_L\},j\in\{1,\dots,n_L\}} |\hat{A}_{ij}^{\hat{z}} - A_{ij}| =o_p(1), \quad \textnormal{as} \ n \rightarrow \infty, T \rightarrow \infty.
\end{align*}
\end{manualtheorem}

%%%%%%%%%%%%%%============= Section 2
\section{Additional Discussion of the Hub Model with the Null Component and Unknown Hub Set}

% \tcb{Add the disscussion on the necessary conditions of the identifiability.}
We give a new identifiability result for the hub model with the null component and unknown hub set.
Recall that $V_0$ is the true hub set with $|V_0| = n_L$. 
Let $\tilde{V}_0$ be another potential hub set with the corresponding parameters $(\tilde{\rho},\tilde{A}) \in \mathcal{P}$ such that 
$\mathbb{P}(g|\rho,A) = \mathbb{P}(g|\tilde{\rho},\tilde{A})$.

\begin{manualtheorem}{S1}\label{thm:thmS1}
	The parameters $(\rho,A)$ of the hub model with the null component and unknown hub set are identifiable under the following conditions:
	\begin{enumerate}
		\item[($i'$)] $A_{ij} < 1$ for $i\in V_0 \cup \{0\}$ and 
		$\tilde{A}_{ij} < 1$ for $i\in \tilde{V}_0 \cup \{0\}, j=1,\dots,n, j\neq i$;
		\item[($ii'$)] for all $i\in V_0,i^{\prime}\in V_0, i\neq i^{\prime}$, there exists $k \in V \setminus V_0$ such that $A_{ik} \neq A_{i^{\prime}k}$;
		\item[($iii'$)] for all $i\in V_0$, there exist $k,k' \in V \setminus V_0$ and $k\neq k'
		$ such that $\pi_k\neq A_{ik}$ and $\pi_{k^{\prime}}\neq A_{ik^{\prime}}$;
		\item[($iv'$)] there exists $k \notin V_0 \cup \tilde{V}_0$ such that for any $i\in V_0$, $\pi_k\neq A_{ik}$, and for any $l \in \tilde{V}_0$, $\tilde{\pi}_{k} \neq \tilde{A}_{lk}$.
	\end{enumerate}
\end{manualtheorem}
Conditions (i') - (iii') are identical to those in Theorem 1 and Theorem 2.
Condition (iv') requires there exists at least one node that can only play a role as a follower.

\begin{proof}[Proof of Theorem \ref{thm:thmS1}]

%We show the following claim  by contradiction: there does not exist a different set of parameters $(\tilde{\rho},\tilde{A})$ such that  $\mathbb{P}(g|{\rho},{A}) = \mathbb{P}(g|\tilde{\rho},\tilde{A})$ for any $g$. 
Theorem 2 shows when $V_0 = \tilde{V}_0$, the parameters in the hub model with null component are identifiable. Therefore, we only need to show $V_0 = \tilde{V}_0$ if $\mathbb{P}(g|{\rho},{A}) = \mathbb{P}(g|\tilde{\rho},\tilde{A})$ for all $g$.

Suppose there exist $(\tilde{\rho},\tilde{A}) \neq (\rho,A) $ such that $\mathbb{P}(g|\rho,A) = \mathbb{P}(g|\tilde{\rho},\tilde{A})$ for any $g$.
Let $B_1 = \tilde{V}_0\setminus V_0$ and $B_2 = V \setminus (V_0 \cup \tilde{V}_0)$.
First, we consider the probability that no node appears   
\begin{equation}\label{eq:none1}
\rho_0\prod_{j=1}^n (1-A_{0j}) = \tilde{\rho}_0\prod_{j=1}^n (1-\tilde{A}_{0j}),
\end{equation}
and the probability that only $k\in B_2$ appears,
\begin{equation}\label{eq:k1}
\rho_0 A_{0k}\prod_{j\neq k}^n (1-A_{0j}) = \tilde{\rho}_0\tilde{A}_{0k}\prod_{j\neq k}^n (1-\tilde{A}_{0j}).
\end{equation}
Dividing \eqref{eq:k1} by \eqref{eq:none1}, since $A_{0k}<1$, we have $A_{0k} = \tilde{A}_{0k}$ for any $k\in B_2$.

Next we show that $B_1 = \tilde{V}_0\setminus V_0 = \emptyset$. 
Suppose $B_1\neq \emptyset$. By condition (\romannumeral 4'), for any $i \in B_1$, there exists a $k\in B_2$ such that $\tilde{A}_{0k} \neq \tilde{A}_{ik}$. Consider the probability that only $i$ appears,
\begin{equation}\label{eq:c1}
\tilde{\rho}_0\tilde{A}_{0i}\prod_{j=1,\dots,n,j\neq i}(1-\tilde{A}_{0j}) + \tilde{\rho} _i\prod_{j=1,\dots,n,j\neq i} (1-\tilde{A}_{ij}) ={\rho}_0{A}_{0i}(1-A_{0k})\prod_{j=1,\dots,n,j\notin \{i,k\}}(1-{A}_{0j}) ,
\end{equation}
and the probability that only $i$ and $k$ appear
\begin{equation}\label{eq:ck1}
\tilde{\rho}_0\tilde{A}_{0i}{A}_{0k}\prod_{j=1,\dots,n,j\notin \{i,k\}}(1-\tilde{A}_{0j}) + \tilde{\rho}_i\tilde{A}_{ik}\prod_{j\notin \{i,k\}}(1-\tilde{A}_{ij}) ={\rho}_0{A}_{0i}A_{0k}\prod_{j\notin \{i,k\}}(1-{A}_{0j}). 
\end{equation}
	
Let 
\begin{align*}
&\tilde{x} = \tilde{\rho}_0\tilde{A}_{0i}\prod_{j=1,\dots,n,j\notin \{i,k\}}(1-\tilde{A}_{0j}),\\
&\tilde{y} = \tilde{\rho}_i\prod_{j=1,\dots,n,j\notin \{i,k\}} (1-\tilde{A}_{ij}).
\end{align*}
Then \eqref{eq:c1} and \eqref{eq:ck1} can be viewed as a system of linear equations with unknown variables $\tilde{x}$ and $\tilde{y}$:
$$
\begin{pmatrix}
  {A}_{0k} & \tilde{A}_{ik} \\
  1-{A}_{0k} & 1-\tilde{A}_{ik}	
\end{pmatrix}
\begin{pmatrix}
  \tilde{x}\\
  \tilde{y}
\end{pmatrix}
=
\begin{pmatrix}
  {\rho}_0{A}_{0i}A_{0k}\prod_{j=1,\dots,n,j\notin \{i,k\}}(1-{A}_{0j}) \\
  {\rho}_0{A}_{0i}(1-A_{0k})\prod_{j=1,\dots,n,j\notin \{i,k\}}(1-{A}_{0j})
\end{pmatrix}.
$$
Since ${A}_{0k} = \tilde{A}_{0k} \neq \tilde{A}_{ik}$, the system is full rank and hence has a unique solution:
\begin{align*}
\tilde{\rho}_0\tilde{A}_{0i}\prod_{j=1,\dots,n,j\notin \{i,k\}}(1-\tilde{A}_{0j}) &= {\rho}_0{A}_{0i}\prod_{j=1,\dots,n,j\notin \{i,k\}}(1-{A}_{0j}),\\
\tilde{\rho}_i\prod_{j=1,\dots,n,j\notin \{i,k\}}(1-\tilde{A}_{ij}) &= 0.
\end{align*}
Combining with \eqref{eq:none1}, we have
\begin{align*}
\tilde{\rho}_0 (1-\tilde{A}_{0i}) \prod_{j=1,\dots,n,j\notin \{i,k\}}(1-\tilde{A}_{0j}) &=  {\rho}_0(1-{A}_{0i})\prod_{j=1,\dots,n,j\notin \{i,k\}}(1-{A}_{0j}).
\end{align*}
As $\tilde{A}_{ij}< 1$, $A_{0i} = \tilde{A}_{0i}$ for any $i \in B_1\subset \tilde{V}_0$ and $\tilde{\rho}_i=0$, which  contradicts the assumption that $0<\tilde{\rho}_i<1$ for any $i \in \tilde{V}_0$. Therefore, $\tilde{V}_0\setminus V_0 = \emptyset$ implies that $\tilde{V}_0$ does not contain any redundant component.
	
By the same argument, we obtain $A_{0i} = \tilde{A}_{0i}$ for any $i \in V_0 \setminus \tilde{V}_0$ and ${\rho}_i=0$, which  contradicts  the assumption $0<{\rho}_i<1$ for $i \in {V}_0$.
Therefore, $V_0 \setminus \tilde{V}_0= \emptyset$. Hence, $V_0 = \tilde{V}_0$. By Theorem 2, we have $(\tilde{\rho},\tilde{A}) = (\rho,A)$.
\end{proof}

We close this section by a discussion on how the penalized log-likelihood function (Section 3.2 in the main text) can result in sparse solutions. Maximizing the Lagrangian form of the penalized log-likelihood function is equivalent to maximizing $L(A,\rho)$ under the following constraints
\begin{equation*}
	\begin{split}
	 \rho_i \geq 0, \,\, i=0,1,\dots,M, \,\, \sum_{i=0}^M \rho_i = 1,\,\,	  \sum_{i=1}^{M}[\log (\epsilon + \rho_i)-\log \epsilon] \leq t.
	\end{split}
\end{equation*}
To show  how the constraints can result in sparse solutions, we consider a toy model containing only two nodes, both of which are potential hub set members, that is, $M=2$.
The constraints become
\begin{align}
	& \rho_1 \geq 0, \,\, \rho_2 \geq 0, \,\, \rho_1 + \rho_2  \leq 1, \label{default_region} \\
	& \log(1+\frac{\rho_1}{\epsilon}) + \log(1+\frac{\rho_2}{\epsilon}) \leq t. \nonumber 
\end{align}
\begin{figure}[!htp]
\centering
\includegraphics[width=0.5\textwidth]{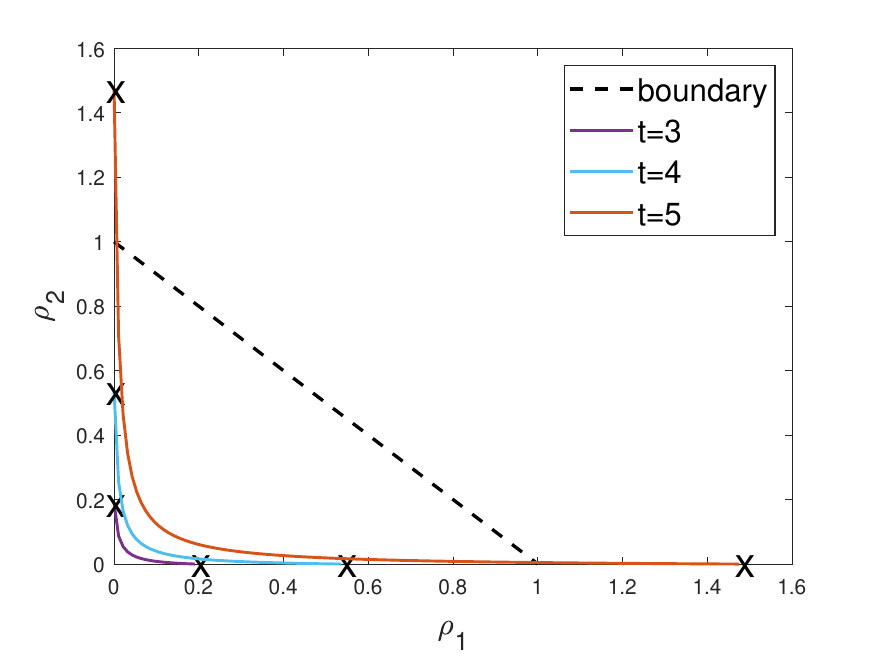}
\caption{Feasible regions of the log penalty with different  values of $t$.}
\label{fig:lp}
\end{figure}
%\begin{figure}[ht!]
%	\begin{center}
%		\twoImages{fig//log.png}{5cm}{log penalities} {fig//lp.png}{5cm}{$L^p$ penalties}
%	\end{center}
%	\caption{two different type of penalties: $L^p$ norms vs. log penalties}%
%	\label{fig:lp}
%\end{figure}

%%%%%%%%%%%%%%%%%%==========================Section 3
Figure \ref{fig:lp} shows the feasible regions of the log penalties for $t=3,4,5$ and $\epsilon = 0.01$, where the crosses mark the intersection of $\log(1+\rho_1/{\epsilon}) + \log(1+{\rho_2}/{\epsilon}) = t$ and the axes, and the dashed line indicates $\rho_1+\rho_2=1$. For $t=3$ and $4$, $\hat{\rho}_1$ (resp. $\hat{\rho}_2$) can potentially reach 0 with $\hat{\rho}_2$ (resp. $\hat{\rho}_1$) being non-zero, indicated by the cross markers within the  region defined by \eqref{default_region}. For $t=5$ (corresponding to a smaller $\lambda$), this cannot happen because $\log(1+\rho_1/{\epsilon}) + \log(1+{\rho_2}/{\epsilon}) = 5$ intersects with the axes outside of the  region defined by \eqref{default_region}. 

\section{Additional Simulation Results}

To further study the performance of the estimates under the setting of sparse $A$, we introduce a scale factor $\alpha$ to control the density of $A$.
Specifically, $A_{ij} \sim U(0.2\alpha,0.4\alpha)$ for $j \in V_i$ and $A_{ij} \sim U(0,0.2\alpha)$ for $j \notin V_i$, where $\alpha = 0.1,0.2,\dots,1$.
We study how the ratios of the RMSEs when the hub labels are unknown to those when the hub labels are known i.e., $\textnormal {RMSE}(\hat{A}_{ij})/\textnormal{RMSE}^*$, change with the degree of sparsity.  We  present the results for the case when $n = 100$. Other simulation settings are the same with those in Section 4.1.

\begin{figure}[!htp]
    \centering
    \includegraphics[width=0.9\textwidth]{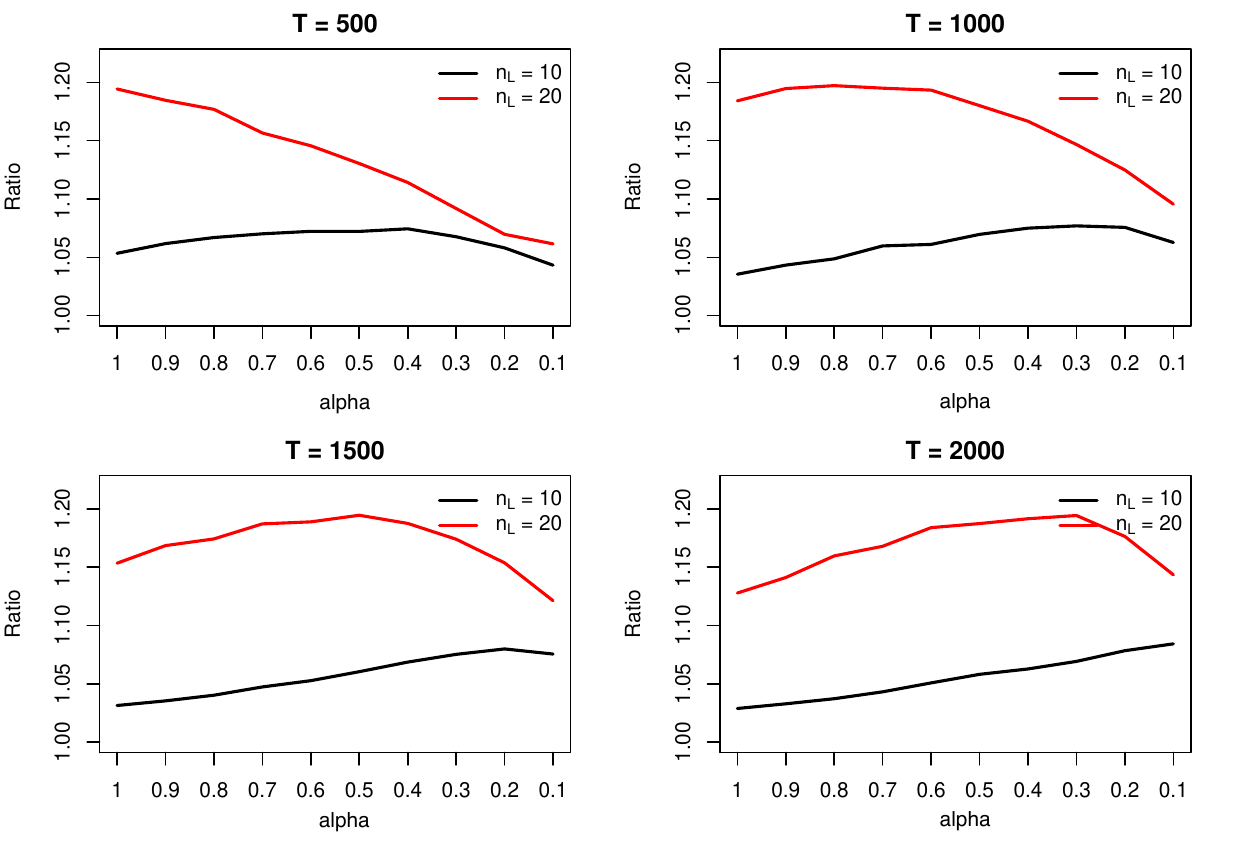}
    \caption{The asymmetric hub model results. The ratio is $\textnormal{RMSE}(\hat{A}_{ij})/\textnormal{RMSE}^*$.}
    \label{fig:asymmetric_hub}
\end{figure}

\begin{figure}[!htp]
    \centering
    \includegraphics[width=0.9\textwidth]{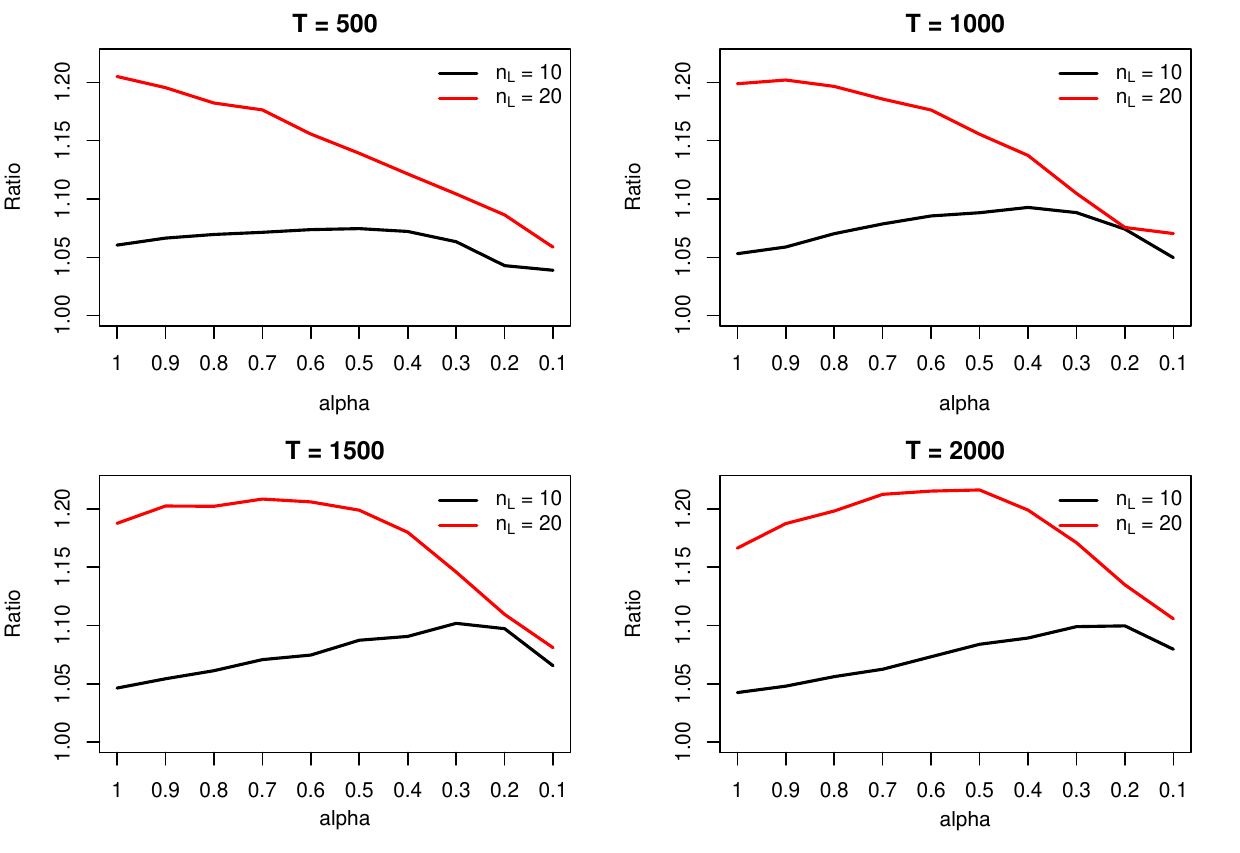}
    \caption{The hub model with the null component results. The ratio is $\textnormal{RMSE}(\hat{A}_{ij})/\textnormal{RMSE}^*$.}
    \label{fig:null_hub}
\end{figure}

Figure \ref{fig:asymmetric_hub} and \ref{fig:null_hub} show the results of ratio versus $\alpha$ for the asymmetric hub model and the hub model with the null component, respectively. 
As $\alpha$ decreases, the ratio typically first increases and then decreases. This suggests that the estimators in both cases perform well when $A$ is dense, and the problem becomes more difficult for the estimator with unknown hubs as $A$ becomes sparser. However, when $A$ becomes too sparse, the matrix $A$ cannot be well estimated even for the case of known hub labels (i.e., the baseline). 

Moreover, Figure \ref{fig:asymmetric_hub} and \ref{fig:null_hub} show that the turning point, i.e., the maximizer of the ratio, comes earlier when $A$ is more difficult to estimate, which corresponds to the cases with larger $n_L$, smaller $T$, and the hub model with the null component. The turning point corresponds to the $\alpha$ value that gives the largest gap between the RMSE for the estimator with unknown hub labels and the baseline, and when the settings become more difficult, the estimator with unknown hub labels starts to face challenges on a denser graph.

%of ratio comes earlier (larger $\alpha$) for the cases with larger $n_L$, smaller $T$ and the hub model with the null component.  
%It makes sense because it becomes more difficult to estimate $A$ for these cases.

%%%%%%%%%%%%%%%%%%%%%%==========================Section 4

\section{Additional Analysis of Passerine Data }
We bootstrap 1,000 samples from the original data to evaluate the stability of the proposed hub set selection method. Specifically, we perform our method on each bootstrapped sample under $\lambda$ from 0.045 to 0.065 and compute the proportion of  each node being selected as a hub set member.
Table \ref{tbl:boot} demonstrates the stability of the proposed method: the majority of the birds are not selected as a hub set member in any bootstrap sample, and $v_9$, $v_{30}$ and $v_{42}$, the three birds identified from the original data dominate in the selection proportions across the bootstrapped samples. 

\begin{sidewaystable}
	\centering
	\caption{Selection proportion from bootstrap}
	\small
	\begin{tabular}{cccccccccccccccc}
		\hline
		$\lambda$ & $v_1$ & $v_2$& $v_3$& $v_4$& $v_5$& $v_6$& $v_7$& $v_8$& $v_9$& $v_{10}$& $v_{11}$& $v_{12}$& $v_{13}$& $v_{14}$& $v_{15}$\\
		\hline
		0.045 & 0 & 0 & 0 & 0.045 & 0 & 0 & 0.81 & 0 & 0.995 & 0.870  & 0 & 0 & 0 & 0 & 0\\
		0.050 & 0 & 0 & 0 & 0.050 & 0 & 0 & 0    & 0 & 1     & 0.600  & 0 & 0 & 0 & 0 & 0\\ 
		0.055 & 0 & 0 & 0 & 0     & 0 & 0 & 0    & 0 & 1     & 0      & 0 & 0 & 0 & 0 & 0\\ 
		0.060 & 0 & 0 & 0 & 0.005 & 0 & 0 & 0    & 0 & 0.965 & 0      & 0 & 0 & 0 & 0 & 0\\
		0.065 & 0 & 0 & 0 & 0     & 0 & 0 & 0    & 0 & 0     & 0.005  & 0 & 0 & 0 & 0 & 0 \\ 
		\hline
		$\lambda$& $v_{16}$& $v_{17}$& $v_{18}$& $v_{19}$& $v_{20}$& $v_{21}$& $v_{22}$& $v_{23}$& $v_{24}$& $v_{25}$& $v_{26}$& $v_{27}$& $v_{28}$& $v_{29}$& $v_{30}$\\
		\hline
		0.045 & 0.01 & 0 & 0 & 0 & 0.810 & 0 & 0 & 0 & 0.010 & 0.095 & 0.115 & 0.025 & 0 & 0 & 0.890\\
		0.050 & 0    & 0 & 0 & 0 & 0.600 & 0 & 0 & 0 & 0.015 & 0.075 & 0.100 & 0.015 & 0 & 0 & 0.625\\
		0.055 & 0    & 0 & 0 & 0 & 0.005 & 0 & 0 & 0 & 0     & 0.005 & 0     & 0.005 & 0 & 0 & 0.945\\
		0.060 & 0    & 0 & 0 & 0 & 0.025 & 0 & 0 & 0 & 0     & 0     & 0.015 & 0.005 & 0 & 0 & 0.830\\
		0.065 & 0    & 0 & 0 & 0 & 0.010 & 0 & 0 & 0 & 0     & 0     & 0.005 & 0     & 0 & 0 & 0.015\\
		\hline
		$\lambda$& $v_{31}$& $v_{32}$& $v_{33}$& $v_{34}$& $v_{35}$& $v_{36}$& $v_{37}$& $v_{38}$& $v_{39}$& $v_{40}$& $v_{41}$& $v_{42}$& $v_{43}$& $v_{44}$& $v_{45}$\\
		\hline
		0.045 & 0 & 0 & 0.825 & 0 & 0 & 0 & 0.830 & 0 & 0 & 0 & 0 & 0.965 & 0 & 0.005 & 0\\
		0.050 & 0 & 0 & 0.625 & 0 & 0 & 0 & 0.105 & 0 & 0 & 0 & 0 & 0.935 & 0 & 0     & 0\\
		0.055 & 0 & 0 & 0.010 & 0 & 0 & 0 & 0.015 & 0 & 0 & 0 & 0 & 0.985 & 0 & 0     & 0\\
		0.060 & 0 & 0 & 0.040 & 0 & 0 & 0 & 0.020 & 0 & 0 & 0 & 0 & 0.910 & 0 & 0     & 0\\
		0.065 & 0 & 0 & 0.045 & 0 & 0 & 0 & 0.050 & 0 & 0 & 0 & 0 & 0.080 & 0 & 0     & 0\\
		\hline
		$\lambda$& $v_{46}$& $v_{47}$& $v_{48}$& $v_{49}$& $v_{50}$& $v_{51}$& $v_{52}$& $v_{53}$& $v_{54}$& $v_{55}$& & & & & \\
		\hline
		0.045 & 0.845 & 0 & 0 & 0 & 0 & 0 & 0 & 0 & 0 & 0 &  &  &  &  &\\
		0.050 & 0.235 & 0 & 0 & 0 & 0 & 0 & 0 & 0 & 0 & 0 &  &  &  &  &\\
		0.055 & 0     & 0 & 0 & 0 & 0 & 0 & 0 & 0 & 0 & 0 &  &  &  &  &  \\
		0.060 & 0     & 0 & 0 & 0 & 0 & 0 & 0 & 0 & 0 & 0 &  &  &  &  &  \\
		0.065 & 0     & 0 & 0 & 0 & 0 & 0 & 0 & 0 & 0 & 0 &  &  &  &  &  \\
		\hline	
	\end{tabular}\label{tbl:boot}
\end{sidewaystable}

\section{Analysis of Extended Bakery Data} 

We apply the hub model with the null component to the extended bakery dataset (available at \url{http://wiki.csc.calpoly.edu/datasets/wiki/ExtendedBakery}) to find the hub items and relationships among all the items. 
The dataset is a collection of purchases in a chain of bakery stores.
The stores provide 50 items including 40 bakery goods (1-40) and 10 drinks (41-50).
The goods can be divided into five categories: cakes (1-10), tarts (11-20), cookies (21-30) and pastries (31-40).
Each purchase contains a collection of items bought together.

The extended bakery data was used as a benchmark dataset to test certain machine learning methods. For example,  \cite{agarwal2016association} used association rule mining to extract the hidden  relationships of items and \cite{negahban2018learning} applied a multinomial logit (MNL) model to address the problem of collaboratively learning representations of the users and the items in recommendation systems.

%is also analyzed by some classic methods, such as \cite{agarwal2016association} used Association Rule Mining to extract the hidden strong relationships of items, \cite{negahban2018learning} applied a MultiNomial Logit (MNL) model to address the problem of collaboratively learning representations of the users and the items in recommendation systems.

In our experiment, we use the 5,000 receipts in the dataset.
Since drinks are typically purchased  as affiliated items of food, we use the 40 bakery goods as the potential hub set, i.e., $\bar{V}_0 = \{1,\dots,40\}$. We use  $\lambda =0.025,0.030,\dots,0.045$ to estimate the  hub set. 
\begin{table}[!htp]
\centering
\caption{Estimated hub set for extended bakery data}
\label{data:bakery}
\begin{tabular}{c|ccccccccc}
\hline
 $\lambda$ & \multicolumn{9}{c}{Selected hub nodes}  \\
\hline
0.025  & 1 & 4 & 5 & 6 & 12 & 13 & 25 & 29 & 33 \\
0.030  & 1 & 4 & 5 & 15 & 23 & 29 & 33 & & \\
0.035  & 5 & 15 & 23 & 29 & 34 & & & &\\
0.040  & 15 & 16 & 23 & 29 & 34 & & & &\\
%\rowcolor{Gray}
0.045  & 15 & 23 & 29 & 34 & & & & & \\
\hline
\end{tabular}
\end{table}

Table \ref{data:bakery} shows the estimated hub sets.
As $\lambda$ increases, nodes are removed gradually from the hub set.
According to the BIC criteria, the optimal $\lambda$ is 0.045, at which the estimated hub set contains $v_{15},v_{23},v_{29}$ and $v_{34}$, where $v_{15}$ is tart, $v_{23}$ and $v_{29}$ are cookies, and $v_{34}$ is pastry. 

In addition, if the data was fitted by the hub model without the null component, then the entire node set has to be used as the hub set. In fact, each of the 50 items was purchased individually for at least once, and therefore must serve as a hub if the hubless groups are not assumed. When the hub model with the null component is used, the corresponding items may be removed from the hub set, which greatly reduces the model complexity.

%In addition, this dataset contains 50 different singleton purchases each of which consists of only one goods or drinks.  Hence, if the hub model without null component is applied to fit the data, the hub set will be the population. However, in the hub model with null component, some of the singleton groups, especially those where the singleton appears infrequently in general, are treated as hubless groups. The corresponding nodes may be removed from the hub set, which greatly reduces the model complexity. In this data, the estimated hub set contains only four nodes making the dimension of $A$ changes from $50 \times 50$ to $5\times 50$ (includes the null component).

\clearpage
\bibliography{allref}

\end{document}